\documentclass[manuscript]{aastex}

\slugcomment{ }

\shorttitle{Near-infrared S0 Survey}
\shortauthors{Laurikainen et al.}

\begin{document}


\title{The Near-infrared S0 Survey III: Morphology of 15 Southern Early-type Disk Galaxies}


\author{Eija Laurikainen, Heikki Salo}
\affil{Division of Astronomy, Department of Physical Sciences, University 
of Oulu, FIN-90014, Finland }
\email{eija.laurikainen@oulu.fi}

\author{Ronald Buta}
\affil{Department of Physics and Astronomy, University of Alabama, Box 870324, Tuscaloosa, AL 35487}

\author{Johan Knapen}
\affil{University of Hertfordshire, Centre for Astrophysics Research, Hatfield, Herts AL10, 9AB, UK}

\author{Tom Speltincx}
\affil{Division of Astronomy, Department of Physical Sciences, University
of Oulu, FIN-90014, Finland}

\author{David Block}
\affil{Department of Applied and Computational Mathematics, University of Witwaterstrand,
Johannesburg, South Africa}



\begin{abstract}

Structural analysis has been performed for a sample of 15 Southern
early-type disk galaxies, mainly S0s, using high resolution $K_s$-band images.
The galaxies are mostly barred and many of them show multiple 
structures including bars and ovals, typical for S0s. The new
images are of sufficient quality to reveal new detail on the morphology of 
the galaxies. For example, we report a hitherto undetected nuclear ring 
in NGC 1387, a nuclear bar in NGC 1326, and in the residual image 
also a weak primary bar in NGC 1317.
For the galaxies we (1) measure the radial profiles of the 
orientation parameters derived from the elliptical isophotes, 
(2) apply Fourier methods for calculating tangential forces,
and particularly, (3) apply structural decomposition methods.  
For galaxies with multiple structures a 2-dimensional method is found to be
superior to a 1-dimensional method, but only if in addition to the bulge and the disk,
at least one other component is taken into account. {\it We find
strong evidence of pseudo-bulges in S0s}: ten of the galaxies have
the shape parameter of the bulge near to $n$ = 2, indicating that
the bulges are more disk-like than following the R$^{1/4}$-law.
Also, six of the galaxies have either nuclear rings, nuclear bars
or nuclear disks. In all non-elliptical galaxies in our sample
the $B/T < $ 0.4, as typically found in galaxies having pseudo-bulges.
{\it In two of the galaxies the $B/T$ flux ratio is as
small as in typical Sc-type spirals}. This might be the hitherto 
undiscovered link in the scenario in which spirals are transformed
into S0s.
{\it Also, bars in S0s are found to be shorter and less massive, and have 
smaller bar torques than bars in S0/a-type galaxies}.
\end{abstract}



\section{Introduction}

Structural components of galaxies like bars, bulges, ovals and bar-related
resonance rings are among the key factors for evaluating when present-day 
galaxies formed and how they evolve over time. 
Bulges in early-type disk galaxies are generally assumed to be old,
formed through hierarchical clustering, but they might also be more 
recent structures assembled by internal dynamical processes in galaxies.
In favorable conditions these processes might lead to transformation of spirals into S0s, 
through gas stripping in the halo and the disk (Bekki, Warrick $\&$ Yashuro 2002), or due to
strong and sustained star formation in the central regions of galaxies (Kormendy $\&$ Kennicutt,
hereafter KK2004).

\vskip 0.25cm
In these secular evolutionary processes the relative mass of the bulge increases.
In case of gas stripping
the bulges should have older stellar populations than the disks and their surface
brightness profiles are expected to resemble more the $R^{1/4}$ law than the 
exponential function. 
However,
due to a small amount of gas in S0s, star formation is not expected
to be the dominant mechanism for the growth of the bulge, unless exceptionally
strong accretion of external gas is taking place in these galaxies.
Bulge-like mass concentrations can form also through  
resonant scattering or through bending instabilities in the central regions
of primary bars (Pfenniger 1984; Raha et al. 1991), in which case these so called
``pseudo-bulges'' should have similar stellar populations as the underlying bars.
In order to evaluate the possible role of these processes in galaxy evolution, 
the properties of bars, ovals and bulges need to be systematically 
studied in the spiral/S0 divide, e.g. through Sa, S0/a, to S0$^+$, S0$^-$ types.

\vskip 0.25cm
Secondary processes in galaxy evolution are generally tied
to the growth of a bulge in one way or another, but an
alternative approach was also discussed by van den Bergh
(1979, 1998), who outlined the DDO system of classification
in terms of a nurture theory. In this classification system,
S0s form a sequence parallel to spirals with types S0a,
S0b, and S0c being the S0 analogues of Sa, Sb, and Sc galaxies.
A difficulty with this approach has been that no S0c galaxies 
have been found so far, 
that is, clear S0 galaxies showing a bulge to total
luminosity ratio similar to an Sc galaxy. However, this might be
a bias due to the fact that application of too simple analysis 
methods can easily hide the real nature of bulges, particularly
in S0s which have rich morphological structures.

\vskip 0.25cm

To our knowledge no previous systematic morphological study of 
galaxies has been carried out in the spiral/S0 divide in the near-infrared. Also,
although 2-dimensional decompositions have become a standard, even in  
recent studies (de Souza, Gadotti $\&$ dos Anjos 2004, hereafter SGA2004;
Christlein $\&$ Zabludoff 2004) multiple structures have not been taken into account.
The current study is a third paper in a series where morphological 
properties of a sample of 170 early-type disk galaxies are studied.
The description of the sample and the first results are presented
in Paper I by Laurikainen, Salo $\&$ Buta (hereafter LSB2005), and
in Paper II by Buta et al. (2006, in press). In paper I we
found, for a subsample of 24 S0-S0/a galaxies, that the $B/T$ flux ratios 
are significantly smaller, and the bulges more 
disk-like, than generally assumed for their morphological type.
In paper II we investigate in detail the Fourier
properties of early-type bars and ovals, and identify bars, which are well described    
in terms of single or double Gaussian components of the relative
Fourier amplitudes of density.
In this study we analyze the morphological 
properties of 15 Southern galaxies of the de Vaucouleurs revised Hubble types S0$^-$ to Sa, 
mainly S0s, with a large range of apparent bar strengths.
For these galaxies we use $K_s$-band images to derive 
the orientation parameters from the elliptical isophotes, apply 
Fourier methods for calculating non-axisymmetric forces, and perform 2D multicomponent 
structural decompositions. The multicomponent decompositions are compared with 
more simple bulge/disk decompositions and with 1D decompositions for the same galaxies.
 
\vskip 0.25cm

\section{Sample and data reductions }
\vskip 0.25cm
This is a sub-sample of 15 galaxies of the Near-InfraRed S0-Survey
(NIRS0S), which was originally drawn from the Third Reference Catalog
of Bright Galaxies (de Vaucouleurs et al. 1991, RC3) 
having the following selection criteria: 
-3 $\leq$ $T$ $<$ 2, $B_T$ $<$ 12.5 mag and logR$_{25}$ $<$ 0.35, 
yielding a sample of 170 galaxies. The final sample consists of
190 galaxies, because 20 late-type ellipticals classified as 
S0s by Sandage, were also added. This is
for taking into account also the misclassified Es in RC3.
The original sample was selected to be similar 
in size with the Ohio State University Bright Galaxy Survey (OSUBGS;
(Eskridge et al. 2002) for spirals. The properties of the sample 
are described in more detail 
in Paper II.
\vskip 0.25cm

The observations were carried out in December 2004 using the SOFI instrument, attached
to the 3.5 m New Technology Telescope (NTT) at ESO. High resolution $K_s$-band images 
(0.29 arcsec/pix) were obtained for all galaxies, and for five of them
also $J$-band images. The total on-source
integration time was 1800-2400 sec in the $K_s$-band, and 1200 
sec in the $J$-band, mostly taken in snap-shots of 3-6 seconds. Owing to the
high sky brightness in the near-infrared, the sky fields were 
taken alternating between the galaxy and the sky field in time intervals
of 1 min: in the object a random dithering within a box of 20 arcsec was used, whereas
the sky fields were randomly selected carefully avoiding bright 
foreground stars. The observations were obtained in good
atmospheric conditions, the nights being mainly photometric. The mean stellar Full 
Width at Half Maximum (FWHM) is 0.72 arcseconds in the $K_s$-band. The sample and the 
seeing conditions are shown in Table 1, where stdev is the standard deviation
of the mean in the sky background.
The galaxy classifications in column 3 are 
from RC3. The adopted morphological type in column 2 is mainly from The de Vaucouleurs 
Atlas of Galaxies by Buta, Corwin $\&$ Odewahn (2007, hereafter BCO2007), 
or the classification is judged by Buta using the images in Sandage $\&$ Bedke catalog (1994).
For NGC 3706 and ESO 208-G21, the classification is based on
the $K_s$-band images from this study. 
\vskip 0.25cm

The images were processed using IRAF\footnote{IRAF is distributed
by the National Optical Astronomy Observatories, which are operated
by AURA, Inc., under cooperative agreement with the National Science
Foundation.} package routines. 
The reduction steps consisted of sky subtraction, flat-fielding of the individual
difference images, removal of foreground stars, and transposing the images
to have north up and west on the right. Dome flats were used, in which the
shade patterns due to the scattered light were corrected. The uneven
illumination of the screen was also corrected in the flat-fields using a 
correction frame, created from multiple observations of one standard star.
The standard star observation was used to create a surface, which was 
then normalized to unity. For these corrections
IRAF scripts offered by ESO were used.  
The images were cleaned of foreground stars by finding
the stars with DAOFIND, and then using a stellar PSF to remove
the stars. Any residuals left by the PSF fitting were cleaned 
using IRAF routine IMEDIT. For five of the galaxies $J-K_s$
color maps were also constructed. The images and the
color maps are shown in Figure. 1. All direct images in Figure 1
are in units of mag arcsec$^{-2}$ and are generally displayed from
$\mu_{K_s}$ = 12.0 to 22.0 mag arcsec$^{-2}$. Color index maps display 
colors in the range 0.5 $<$ $J-K_s$ $<$ 1.5. 
\vskip 0.25cm

\section{Elliptical isophotes}
\vskip 0.25cm

The elliptical isophotes were derived mainly in the $K_s$-band using the
ELLIPSE routine in IRAF. We calculate the radial profiles of the position angles $\phi$ 
and the ellipticity $\epsilon$ = 1 - $q$, where $q$ is the 
minor-to-major axis ratio (see Fig. 2). The profiles were calculated
both in linear and in logarithmic scale: the linear scale 
for deriving the orientation parameters of the
outer disks, and the logarithmic scale for identifying the structural components
particularly in the inner regions of the galaxies. 
The isophotal fits were made to the original images. 
When available, $J$-band images were used to estimate the orientation
parameters, because they were generally deeper than the $K_s$-band images.
In the isophotal fitting deviant pixels above $3 \ \sigma$ were rejected. 
The parameters $\phi$ and $q$ are listed in Table 2, calculated
as mean values in the radial range indicated in the table.
The uncertainties are
standard deviations of the mean in the same radial range.
Also shown are the orientation parameters 
given in RC3, and for some of the galaxies values from other sources as well.
The values used in this study are shown in boldface.
In the analysis orientation parameters derived in this study are
used, except for the galaxies NGC 3081, NGC 1512 and NGC 1326. For 
the first two galaxies we use kinematical estimates of $\phi$ obtained by
Buta $\&$ Purcell (1998) and Buta (1988), respectively, and for
NGC 1326 the photometrical estimates by Buta et al. (1988) is used.
\vskip 0.25cm

A comparison to RC3 shows large deviations in the orientation parameters 
particularly for 
the galaxies NGC 1079 ($\phi$), NGC 1317 ($q$ and $\phi$),
NGC 1512 ($\phi$), NGC 1533 ($q$ and $\phi$) and NGC 3081
($q$, $\phi$). However, our measurements are generally in good 
agreement with those measurements given in the literature, where deep optical or IR images are used. This
is the case for NGC 1512, for which Kuchinsky et al. (1998) 
obtained $\phi$=56$^{\circ}$. For NGC 3081 Buta $\&$ Purcell (1998)
give photometrically estimated values of $\phi$=86$^{\circ}$ and $q$=0.387. Also, for
NGC 1351 our measurements for the orientation parameters are in
agreement with those obtained by Garcia-Gomez et al. (2004).
\vskip 0.25cm

\section{Fourier decompositions}
\vskip 0.25cm

For estimating tangential forces induced by non-axisymmetric structures 
in galaxies we use a gravitational bar torque approach (Combes $\&$ Sanders 1981; 
Quillen, Frogel $\&$ Gonzalez 1994),
first applied to galaxy samples by (Buta $\&$ Block 2001; Laurikainen $\&$ 2002).
The polar method we use has been
described by Laurikainen $\&$ Salo (2002), Salo et al. (2004), and Laurikainen,
Salo $\&$ Buta (2004, hereafter LSB2004) (see also Salo et al. 1999). The gravitational 
potential is inferred from the two-dimensional $K_s$-band image by assuming
that the near-IR light distribution traces the mass and that
the mass-to-luminosity flux ratio is constant. It is assumed that the
vertical profile is exponential and that the vertical scale height
depends on the morphological type. Deprojected images were used, which
were constructed by first subtracting off the bulge model, deprojecting 
only the disk light, and then adding back the bulge as an assumed spherical component
(see Section 5 also for more sophisticated decompositions). 
The derived gravitational 
potentials are used to calculate maps of the maximum ratio
of the tangential force ($F_T$) to the azimuthally averaged radial force
($F_R$), as well as a radial profile of the maximum relative tangential 
force at each distance (see Fig. 3), 

 
\begin{equation}
Q_T(r) = {|F_T(r,\phi)|_{max} \over <|F_R(r,\phi)|>},
\end{equation}

\noindent where $<|F_R(r,\phi)|>$ denotes the azimuthally averaged axisymmetric
force for each $r$. The maximum value of this profile, $Q_g$ at a radius $r_{Qg}$, 
gives a single measure of {\it bar strength} for these early-type
galaxies where spiral arm torques are generally negligible. In the calculation of the gravitational
potential we use azimuthal Fourier decompositions in radial annulae, typically
using even components with azimuthal wave number $m < $ 20. The gravitational 
potential is then calculated using FFT in the integration over azimuth, and direct
summation in radial coordinates. An important advantage of this method,
compared to the more straightforward Cartesian integration (``Quillen's method'') is the suppression
of spurious force maxima in the outer disk arising due to noise. The other
advantage is that it gives simultaneously 
also the relative mass of the bar, as estimated from the maxima of $m$ = 2 and $m$ = 4 
amplitudes of density ($A_2$ and $A_4$, respectively). 
The measurements are shown in Table 3, where the bar lengths are also given.
Bar length is identified within the region where the phase of the $m$ = 2 amplitude
is maintained nearly constant. If no bar length is given,
the parameters indicate the properties induced by other non-axisymmetric components
like ovals or inner disks. The errors are formal errors related to $Q_g$ measurements
in the four image quadrants.
It is worth noticing that a peak in a 
$Q_T$-profile should not be automatically interpreted to indicate a presence of a bar.
For example, in NGC 3100 strong tangential forces are detected at $r$ $<$ 40'',
although there is no bar in this galaxy. As discussed in Section 5,
this galaxy is dominated by a large oval, which appears as a non-axisymmetric
structure in the Fourier analysis.
Because of the high uncertainty 
in the position angle of NGC 1512, no $Q_g$ value for this galaxy is given in Table 3:
the kinematic estimate by Buta (1988) gives $\phi$ = 83$^{\circ}$ while we measure $\phi$ = 46$^{\circ}$.

\vskip 0.25cm
\section{Structural decompositions}
\vskip 0.25cm
   \subsection{The algorithm}  
\vskip 0.25cm
   
We use a 2-dimensional (2D) multicomponent decomposition code, originally presented by 
LSB2005 (see also LSB2004), where the code was also tested.  
In this study we use also a one-dimensional (1D) version of the code  
(Laurikainen $\&$ Salo 2000), which can be efficiently used in conjunction 
with the 2D code.

In the 2D multicomponent code all components use generalized elliptical 
isophotes (Atthanassoula et al. 1990), the isophotal radius being described by the function:


\begin{equation}
r=(|x|^{c+2} + |y/q|^{c+2})^{1/(c+2)}.
\end{equation}

\noindent The isophote is boxy when the shape parameter $c > $ 0, disky when $c < $ 0, and purely
elliptical when $c$ = 0. Circular isophotes correspond to $c$ = 0 and 
minor-to-major axial ratio $q$ = 1.

The disks are described by an exponential function:

\begin{equation}
I_d(r_d) \ = \ I_{0d}  \exp[-(r_d/h_r)],
\end{equation}

\noindent where $I_{0d}$ is the central surface density of the disk,
and $h_r$ is the radial scale length of the disk. The radius $r_d$
is calculated along the disk plane defined by the assumed $\phi$ and $q$ 
(for disk c = 0).
The surface brightness profiles of the bulges are described by a 
generalized Sersic's function:

\begin{equation}
I_b(r_b) \ = \ I_{0b} \exp[-(r_b/h_b)^{\beta}],
\end{equation}

\noindent where $I_{0b}$ is the central surface density, $h_b$ is the
scale parameter of the bulge, and $\beta$ = $1/n$ determines the slope
of the projected surface brightness distribution of the bulge ($n$ = 1 for 
exponential, and $n$ = 4 for the $R^{1/4}$-law). The
parameter $r_b$ is the isophotal radius defined via
the parameters $q_b$, $c_b$ and
$\phi_b$, where the subscript stands for the bulge. In the case of
an elliptical bulge, $\phi_b$ is its major axis position angle measured
counter clockwise from North in the sky plane. Sersic's function
can be applied also for the other non-axisymmetric components, 
or alternatively the Ferrers function is used, described as:
\vskip 0.25cm

%
\begin{eqnarray}
I_{bar}(r_{bar}) \ &=& \ I_{0bar} \ (1-(r_{bar}/a_{bar})^2)^{n_{bar}+0.5}, \ \ r_{bar}<a_{bar}\\
  &=&  0, \ \ \ r_{bar}>a_{bar},
\end{eqnarray}

\noindent where $I_{0bar}$ is the central surface brightness of the
bar, $a_{bar}$ is the bar major axis length, and $n_{bar}$ is the exponent
of the bar model defining the shape of the bar radial profile.  The
isophotal radius ($r_{bar}) $ is defined via parameters $q_{bar}$, $c_{bar}$,
$\phi_{bar}$. As in the 2D version, also in the 1D version 
the bulges are described by a Sersic's function.
\vskip 0.25cm

   \subsection{The fitting procedure} 
\vskip 0.25cm

The fitting was made in flux units in the plane of the sky.
Depending on the choice
of the weighting function more weight can be assigned to the pixels
in the central parts of the galaxies where the flux level is high, or to the
outer regions, where the number of pixels is high. In this study 
we generally use $w_i = 1 / F$, where $F$ is the model flux in pixel $i$.
However, it was shown by LSB2005 that
our method is not sensitive to the choice of the weighting function.
In order to account for the effects of seeing, the model is convolved
with a Gaussian PSF using a FWHM measured for each image. 
\vskip 0.25cm

Due to the large number of free parameters in the 2D approach the fitting is time consuming.
In order to speed the process a good fit was first searched for 
the images rebinned by a factor of 4, which results were then used as 
initial parameters
for more sophisticated decompositions performed for the original images. 
Generally the decompositions were started by finding $h_r$. 
In cases where the presence of an extended disk was 
not immediately obvious, spiral arms or rings were used to confirm   
the presence of the underlying disk. In some cases they were visible only
after subtracting the bulge or disk model. A few images 
were too noisy for finding $h_r$ in a reliable 
manner with the 2D code, in which case the 1D code was used 
by fitting an exponential to the outer parts of the disk.

\vskip 0.25cm
Three different kinds of 2D decompositions were made: 
(1) {\it multicomponent} ($N_{max}$ = 7), (2) {\it 3-component} (typically bulge/disk/bar) 
and (3) {\it 2-component} (bulge/disk) decompositions. Additionally,
{\it simple 3-component} decompositions were made by assuming that the
bulges have spherical light distributions.
Such decompositions are required for deprojecting the images, because it
is difficult to make a 3-dimensional bulge model allowing also flattening
or triaxiality for the bulge. As not all bulges are spherical, some
caution is needed while interpreting the innermost structures of the
deprojected images. The best-fitting multicomponent decompositions
are compiled in Table 4 and illustrated in Figure 4. 
The numbers 1, 2 and 3 in the title line in the table generally indicate the primary, 
secondary and tertiary components.
The component 1 is in most cases the primary bar, but in non-barred
galaxies it can be an oval or an inner disk. The component 2 generally refers to 
the secondary bar or oval, and the component 3 to a large tertiary bar or oval. 
The bulge-to-total ($B/T$) flux ratio can be calculated either using the 
disk-component extrapolated to infinity, or limiting to the
fitted region alone: in this study the non-extrapolated disk is used.  
\vskip 0.25cm

To be sure that we were fitting 
real structures, a preliminary identification of bars, ovals
and disks was made by inspecting (1) the original images and the
$J-K$ color maps, 2) the radial profiles of the orientation parameters,
and 3) the profiles of the Fourier amplitudes of density. In some cases 
also the bulge- or disk model-subtracted images were inspected after-wards.
The non-axisymmetric structures appear
as bumps in the $\epsilon$-profiles, provided that their surface brightnesses 
are high enough compared to that of the underlying disk. 
Also, the position angles show isophotal twists in the bar regions.
The even Fourier amplitudes of density are particularly sensitive for detecting the
non-axisymmetric structures in galaxies.
The identifications of the morphological components and the ring dimensions
are shown in Tables 5 and 6. As in Table 4, also in Table 5 the subscripts 1 and 2
refer to the primary and secondary bars/ovals. Following KK2004
we adopt the terminology that
the flat inner structure is called a ``lens'' if the morphological type
is S0-S0/a, and an ``oval'' when the galaxy has a later morphological type.
A lens has been originally defined by Kormendy (1979) as a feature 
showing a shallow brightness gradient interior to a 
sharp edge, whereas an oval was defined as having a broad elongation in the
light distribution. It seems that in some cases the division according to
the morphological type is artificial. S0s can actually have both ovals and lenses.
However, as the discrimination between ovals and lenses is not straightforward, we 
keep the present terminology.

\vskip 0.25cm

   \subsection{Discussion of individual galaxies} 

\vskip 0.25cm 
In the following we discuss the morphological
properties of the individual galaxies, which in early-type disk galaxies are generally
rich. Detecting these structures depends not only on the wavelength, but 
also on the image resolution and the integration time used. We use deep, high resolution
near-IR images, but in spite of that it is sometimes difficult to identify
the structural components, particularly the bulges.
In order to illustrate this also some details of the decompositions 
are given. In most cases it was not possible to use any
semi-automatic procedure to find reliable solutions.

\vskip 0.25cm
{\bf ESO 208-G21}: ESO 208-G21 is a bulge-dominated system with an elongated 
inner structure. Most probably for this reason, (Corwin, de Vaucouleurs $\&$ de
Vaucouleurs 1985)  
classified it as E4, whereas 
in RC3 the inner structure was interpreted to be
a weak bar/oval, leading to the classification of SAB0$^-$. Martel (2004) 
found that the inner structure is a dusty inner disk, which interpretation
was based on the analysis of high resolution $HST$-images.
The disk is clearly visible in our unsharp masked image (Fig. 5) and it can 
be identified also in the ellipticity profile as a peak, followed 
by a decline up to the radius where the dominance of the bulge makes the 
profile flat. The ellipticity profile we find is very similar to that obtained 
previously by Scorza et al. (1998), who also made a structural decomposition for this galaxy.
As they use a different method than we do it is interesting to compare
the results: in both cases a solution was found in which 
the galaxy has a small exponential inner disk, embedded in a large massive bulge. 
The $B/T$ = 0.97 we found is very similar to $B/T$ = 0.92 obtained 
by Scorza et al. in the $V$-band (they give $D/B$ = 0.09). 
The decomposition by Scorza et al.
is based on an iterative process where the disk is first subtracted 
and the isophotes of the remaining bulge component are inspected. The procedure
is repeated until perfect elliptical isophotes remain.  
It is worth noticing that for this galaxy our 1D and 2D codes give very similar 
solutions. Obviously this galaxy has no outer disk, and
the bulge is strongly flattened. We classify this galaxy as
E(d)5, thus confirming the earlier classification by Kormendy $\&$ Bender (1996).

\vskip 0.25cm
{\bf NGC 1079}: The morphology of NGC 1079 has been previously studied 
by Buta (1995),Jungwiert, Combes $\&$ Axon (1997) , SGA2004, who also made a 2D decomposition for it,
and by Knapen et al. (2006), who found a strong nuclear peak in
H$\alpha$ surrounded by rather faint circumnuclear patchy emission.
Buta (1995) recognized NGC 1079 as a galaxy showing one of the largest 
examples of a feature called R$_1$R$_2^{\prime}$, a double outer
ring/pseudoring morphology linked to an outer Lindblad resonance (Buta $\&$ Crocker 1991).
This feature includes a well-defined outer ring from which two conspicuous
spiral arms emerge. Sandage $\&$ Brucato (1979) also recognized 
NGC 1079 as an early-type spiral. The bar in NGC 1079 was recognized by
Jungwiert, Combes $\&$ Axon  as a large peak at $r$ = 30 arcsec
in the ellipticity profile, while a smaller peak at $r$ = 17 arcsec
was attributed to a triaxial bulge. We see both peaks in our $\epsilon$-profile as well.
The axis ratio of the disk ($q$ = 0.59) deviates from the axis ratio of the bulge
($q$ = 0.68), confirming that they are separate components.

\vskip 0.25cm

In our decomposition we first determined $h_r$, which was then kept fixed. Finally the bulge, disk and
the bar were fitted, but for the ansae-type morphology of the bar the solution 
depends slightly on the parameters of the bar: it is possible to use
$n > $ 2, or $n$ = 2 together with the shape parameter that deviates from
purely elliptical isophotes. Depending on this choice we found $B/T$ = 0.20 - 0.25. 
Our solution converges only if flattening of the bulge is taken into account. 
The bar has a high surface brightness compared to that
of the underlying disk. Apparently an inner ring surrounds the bar, seen as a 
residual after subtracting the model image from the observed image.

\vskip 0.25cm

Our solution for this galaxy is very different from the 2D bulge/disk solution 
obtained by SGA2004. They found a large de Vaucouleurs type bulge 
($B/T$ = 0.65, $n$ = 3.82) dominating the whole surface brightness profile, and a 
very weak exponential inner disk, extending to the radius of the bar. In their fit 
the disk at the edge of the bar is about 5 mag weaker than the bulge. 
SGA2004 made the decomposition in the $R$-band, which extends approximately to 
the same radial distance as our $K_s$-band image. For comparison we made 
also a 2D bulge/disk decomposition (thus omitting the bar), 
which gives $B/T$ = 0.57 (see Fig. 6, right upper panel). This
is near the $B/T$-ratio obtained by SGA2004, except that the 
bulge is more exponential ($n$ = 2.7). However, we stress that in this
solution the flux of the bar goes erroneously to the bulge.
A similar solution was found by us using the 1D-method. This resembles the
solution by SGA2004 even more, although the disk we 
find is still several magnitudes brighter than that obtained by SGA2004.
It is also possible to find a solution in which the
surface brightness profile is fitted by a single Sersic's function
with $n$ = 3.3 (Fig. 6, middle upper panel), but in that case the fit 
is not good, and particularly, the model image is very different from
the original image.

\vskip 0.25cm
{\bf NGC 1317}: NGC 1317 has been previously reported to be a double barred
galaxy, both in the optical (Schweizer 1980;  Wozniak et al. 1995), and
in the near-IR (Mulchaey, Regan $\&$ Kundu(1997). The galaxy has also
a nuclear ring, which is actually a double ring/pseudoring evident 
particularly in blue light (Schweizer 1980; Papovich et el. 2003) and in H$\alpha$ 
(Crocker, Baugus $\&$ Buta 1996). The galaxy is nearly in face-on orientation.
For the disk RC3 gives a larger ellipticity than found in this study, 
most probably because the outer isophotes
are elongated towards the nearby companion.
\vskip 0.25cm

The secondary bar is well visible in our $K_s$-band image, and the double nuclear ring
is now detected for the first time also in the near-IR. The outer structure is
dominated by a large oval with no clear sign of a primary bar. However, after
subtracting the exponential disk a very weak classical bar becomes visible (see Fig. 5). 
Previous evidence for the primary bar comes from the $\epsilon$-profiles, which
is largely associated to the large oval.
The secondary bar lies within a bright nuclear lens
(first noticed by Schweizer 1980) and is completely detached from the nuclear
ring/pseudoring. Together, the secondary bar and its lens appear as a bump in the
radial surface brightness profile. In the decomposition only one primary 
component was fitted.
For the complex structure 
of this galaxy the decomposition was made in small steps.
The parameters for the disk and the main bar/oval were found first, after 
which they were fixed. Then the secondary bar was found and 
fixed, and finally the bulge-component was found. In the central part
of the galaxy the secondary bar is dominating over the central oval and
therefore it is also the component that the fitting procedure finds.
In the final solution all components
(except the shape parameter of the bar) were left free for fitting.
The solution we find is reasonable, although it leaves some residuals, mainly
because the bars and ovals cannot be fitted separately.
\vskip 0.25cm

The double nuclear ring in the $J-K_s$ color map shows a presence of dust, 
which is consistent with the observation that the ring is 
presently forming stars. The ring and the secondary bar are also well visible  
in the $K_s$-band image after subtracting the bulge (Fig. 5). The
ring has a patchy morphology, possibly indicating the presence of
red super giants. There are dust lanes also in the central regions of the galaxy,
not aligned with the secondary bar.

\vskip 0.25cm
{\bf NGC 1326}: This is a well studied galaxy previously reported to have
a primary bar, a possible nuclear bar, as well as inner and outer rings. 
The outer ring is an exceptionally bright example of the outer Lindblad 
resonance subclass R$_1$, characterized by an approximate alignment 
perpendicular to the bar and by slight dimpling in its shape near the bar axis 
(Buta $\&$ Crocker 1991). The nuclear ring is also an exceptional 
star-forming feature that has been well studied from both ground
and space (Buta $\&$ Crocker 1991; Garcia 1993; Storchi-Bergman et al. 1996;
Fernandes, Storchi-Bergmann $\&$ Schmitt 1998; Buta et al. 2000)  
The inner bar and the inner and outer rings 
are visible in our $K_s$-band image, and the nuclear ring is evident also in
the $J-K_s$ color map. We detect a faint
secondary bar inside the nuclear ring, which is best visible in 
the $K_s$-band image after subtracting the bulge model (Fig. 5). 
The possible double-barred nature of NGC 1326 was
previously discussed by  de Vaucouleurs (1974), Buta $\&$ Crocker (1991), 
Wozniak et al. (1995) and Erwin (2004), but it has been difficult to verify owing to 
strong extinction inside the nuclear ring. The best evidence presented previously 
was by Erwin (2004), but the resolution in our $K_s$-band image 
is sufficient now to confirm it. The $J-K_s$ color
index map in Figure 1 shows that dust still has some impact in the near-IR bands
inside the nuclear ring. The ring is red in this map which suggests the presence
of red super-giants.
\vskip 0.25cm

In our decomposition for NGC 1326 we fitted besides the bulge and the disk,
also two bars. This solution converges only if flattening of the bulge 
is taken into account. Models
for the individual components were found in small steps and finally 
all components were fitted simultaneously. 
The bulge in this galaxy was found to be quite
prominent ($B/T$ = 0.34, $n$ = 3.0) but  much less luminous than  
argued by SGA2004. They found a very
weak extended disk embedded inside a large de Vaucouleurs type bulge dominating
the whole surface brightness profile, leading to $B/T$ = 0.63 
and $n$ = 4.9. For comparison we made also bulge/disk/bar and bulge/disk
decompositions, the former leading to $B/T$ = 0.58 (and $n$ = 3.1) 
and the latter to $B/T$ = 0.42 (see also Fig. 6).
However, in the bulge/disk decomposition
the flux of the bar/oval goes erroneously  
to the bulge. It is also worth noticing that the B/T in the bulge/disk/bar decomposition
is somewhat larger than in the decomposition where two bars were fitted (B/T = 0.42 v.s. 0.34), 
which means that in this case even the 
bar/oval (visible as a bump at $r$ = 5-10 arcsec in the $\mu$-profile) affects the solution.

\vskip 0.25cm
{\bf NGC 1350}: This galaxy has been previously reported to have a prominent 
inner ring and an outer pseudo-ring  of type R$_1^{\prime}$ (Crocker, Baugus $\&$ Buta 1996). 
In the $K_s$-band NGC 1350 shows a prominent inner ring and an outer
Lindblad resonance subclass R$_1^{\prime}$ pseudoring (Buta $\&$ Crocker 1991).
In the $K_s$-band NGC 1350 shows similar features. The galaxy also 
shows a clear ansae-type bar, both in the direct images and in the
$\epsilon$-profile. The bar is more conspicuous at 2.2 $\mu$m than
in blue light but it is still relatively weak. 
Internal extinction masks much of the bar in blue light. 
There is no clear sign of a secondary bar, but a peak appears
in the $\epsilon$-profile at $r$ = 8 arcsec, associated with an isophotal twist, 
thus at least indicating that a 
separate structural component exists at that radius. It might be an inner
disk, because the orientation deviates only by 13$^{\circ}$ from the orientation
of the outer disk. The bulge has 
a different orientation than the bar so that in our decomposition the two
components cannot be mixed.
\vskip 0.25cm

The different components in the decomposition were found in small steps, and
in the final solution the parameters of the disk and the length of the bar 
were fixed, the latter because for the ansae-type morphology the model would otherwise 
overestimate it. Again, in the robust 2D bulge/disk 
decomposition the flux of the bar goes erroneously to the bulge, giving
$B/T$ = 0.77, instead of $B/T$ = 0.25, obtained in our best fitting solution. 
For this galaxy the flattening of the bulge is high ($q$ = 0.71). 

\vskip 0.25cm
{\bf NGC 1387}: This galaxy is an excellent example of a very early-type
barred galaxy. Although classified as type SAB in RC3, Pahre (1999) and
BCO2007 adopted an SB type. The $Q_b$ family from paper II is
$\underline{\rm A}$B, indicating that the bar is indeed weak.
NGC 1387 has no spiral arms, but the surface brightness 
profile at $r$ = 30 - 90 arcsec is perfectly exponential, which is a strong
sign of a disk. 
\vskip 0.25cm

In the decomposition we fitted, besides the bulge and the disk, also the bar/lens
for which one single function was used. Subtraction of the bulge model
from the original image showed that this galaxy has a prominent nuclear ring at
$r$ = 6 arcsec (Fig. 5), which appears also as a faint peak in the $\epsilon$-profile (Fig. 2).
The ring is nearly circular and has a significant size compared to the length of the bar.
In the surface brightness profile the bulge and the disk 
appear as two nearly exponential functions.

\vskip 0.25cm
{\bf NGC 1411}: Sandage $\&$ Bedke (1994) noticed that NGC 1411 has a three-zone structure with 
a ring between the second and the third components. BCO2007 noted 
a series of lenses, and classified it as type SA(l)0$^{\circ}$, based on a $V$-band
image. In our $K_s$-band image, only the innermost lens clearly stands out, but a weak
outer lens, that has actually the appearance of a smooth ring, 
can also be detected in Figure 5, after subtracting the exponential disk.
The surface brightness profile of this galaxy shows three nearly exponential parts
in the $K$-band image.
\vskip 0.25cm

In spite of the simple looking profile, it was not easy to find a solution 
that converges. Again the 
solution was found in small steps so that robust solutions for the two inner 
components were found first. They were fixed and a solution for the outer 
disk was found. After that the parameters
of the outer disk were fixed and more precise solutions were found for the
two inner components. Finally all parameters of the bulge
and the middle component were left free for fitting. The bulge and the 
middle component were fitted using a Sersic's function, whereas for the outer 
disk an exponential function was used. Our decomposition
confirms the visual impression that all three components are nearly exponential.
We found $B/T$ = 0.08, which is as small as the typical $B/T$-ratios in Sc-type spirals.
Our value for NGC 1411 is not essentially
larger than the $B/T$ = 0.11 found by LSB2005 for Sc-galaxies using the same
decomposition method. A similar small B/T ratio of 0.15 for Sc-galaxies is given also by
Binney $\&$ Merrifield (1998) based on the $R$-band measurements by Kent (1985).
NGC 1411 might therefore be a genuine example
of an S0c galaxy, that is, the proverbial analogue of an Sc galaxy in a
classical S0, as discussed in the 
DDO galaxy classification scheme by van den Bergh (1976, 1998).

\vskip 0.25cm
A more robust 2D bulge/disk decomposition gives $B/T$ = 0.39. In this solution 
the $B/T$ value is very large, because the two 
inner exponentials are fitted by a single Sersic's function. 
For this galaxy the 1D decomposition (see Fig. 6, lower left panel) 
gives a very similar
solution as the 2D bulge/disk decomposition (middle lower panel).
In its simplicity NGC 1411 is an
illustrative example: it clearly demonstrates that if a three-component galaxy is fitted 
by a two-component model, the two-dimensionality alone does not
automatically improve the result. 

\vskip 0.25cm
{\bf NGC 1512}: This is a well-known interacting system and one of 
the latest Hubble types in our sample. The morphology is dominated by 
a strong bar and two rings: a bright oval inner ring around of the bar 
ends and an intense nuclear ring in the center of the bar. 
A deep $B$-band SINGS Legacy image (Kennicutt et al. 2003) also reveals
a distorted outer ring as well as faint tidal debris near the companion
NGC 1510. The nuclear ring
was first noted by Hawarden et al. (1979), who showed also the extent 
of NGC 1512's interaction with neighboring NGC 1510.
Jungwiert, Combes, $\&$ Axon (1997) noticed an isophotal twist at 
$r$ = 7 arcsec, which was suggested to be due to a triaxial 
bulge. However, Erwin (2004) used NICMOS HST images to show that the 
twist is due to the nuclear ring, with no evidence of a bar inside 
the ring (see also de Grijs et al. 2003). The ring (with no bar inside) 
is well visible also in our $K_s$-band image and in the $J-K_s$ color index 
map. As is typical of such rings, the morphology is patchy and 
indicative of strong star formation (Maoz et al. 2001).
\vskip 0.25cm

As our image is not deep enough for a reliable decomposition
for this galaxy, and the position angle is also very uncertain,  
we give only a rough estimate
of the relative mass of the bulge. Only two components are used, a Sersic's function
for the bulge and an exponential function for the bar+disk.
The bulge has nearly circular isophotes and it is fitted by a  
function that approaches an exponential ($n$ = 1.2), thus
indicating a disk-like nature of the bulge. However, we stress that
the decomposition is very uncertain, and therefore no $B/T$ flux ratio 
is given in Table 4.

\vskip 0.25cm
{\bf NGC 1533}: BCO2007 classify 
the galaxy as type (RL)SB0$^{\circ}$, implying a large ring/lens outside the 
bar, but none of the conventional variety, either inner ring or inner lens. 
Sandage $\&$ Bedke (1994) noticed the slight enhancement at the rim 
of the RL and classified the galaxy as an SB0$_2$/SBa. Sandage $\&$ 
Brucato (1979) also noted the large lens under-filled by the bar, and 
they also suspected a weak spiral pattern in the outer lens. 
\vskip 0.25cm

For this galaxy it was possible to fit the bar and the lens separately.
However, it was more difficult to find $h_r$ mainly because the lens 
dominates the disk in our image. Therefore $h_r$ was found 
using the 1D method by fitting an
exponential to the outer disk. In the 2D decomposition $h_r$ 
was fixed to this value.

{\bf NGC 1553}: The dominant feature of this galaxy's morphology 
is a strong inner ring/lens (Sandage $\&$ Brucato 1979; Kormendy 1984). 
The lens shows subtle structure in our $K_s$-band image and is 
classified as an inner ring in RC3. BCO2007 give a type of 
SA($\underline{\rm r}$l)0$^{+}$ to highlight that the 
feature really is more a ring than a lens. In the central part of the 
galaxy, a second elongated feature is present, which in our unsharp mask (Fig. 5)
appears to be a nuclear disk showing also spiral arms.
Sandage $\&$ Bedke (1994) noticed that this 
galaxy has subtle dust lanes in the disk leading to the 
classification of a peculiar SB0/Sa. The major-axis rotation 
curve for this galaxy is found to be typical for S0s (Kormendy 1984; 
Rampazzo 1988). Contrary to all previous evidence, SGA2004 re-classified 
NGC 1553 as an elliptical with 
an inner disk, a choice which was justified by the fact 
that their decomposition did not find any outer disk.
No $B/T$-ratio is given by SGA2004, but the
shape parameter of the bulge was found to be $n$ = 5.24.
\vskip 0.25cm

In our analysis the presence of the inner disk is obvious: it is 
particularly well visible in the image where the exponential disk
is subtracted, shown in Figure 5.
The best 2D solution was found by fitting a bulge, lens/ring and an outer disk, which
appeared to be exponential (as also in the $\mu$-profile by Kormendy 1984). 
We found $B/T$ = 0.21 and $n$ = 1.9.
Again, in order to duplicate the 2D solution
of SGA2004 we also made a 2D bulge/disk
decomposition; this gives $B/T$ = 0.88 and $n$ = 3.33,
which is exactly the same result as obtained also by our 1D method for this galaxy.
The reason for the very large difference between our 2- and 3-component
decompositions is that in the 2-component solution the flux of the
lens goes erroneously to the bulge. 
In all our decompositions the shape parameter of the bulge is considerably smaller
than the value $n$ = 5.24 obtained by SGA2004.
\vskip 0.25cm

The nature of the bulge in this galaxy is unclear even in our best fitting
solution. The component that is fitted as a bulge is actually the inner disk and there
is no other more spherical component in the central regions
that could be interpreted as a bulge.
Perhaps there is no bulge in this galaxy, unless the inner disk is interpreted
as a pseudo-bulge. Kinematical evidence for this interpretation comes
from the study  by Kormendy
(1984) who argued that the bulge (the small central component) is cooler 
than the lens dominating at 
larger radial distances. The change in the stellar velocity dispersion occurs 
at $r$ = 12 -15 arcsec after which the lens starts to dominate. 
This is precisely the distance inside of which we find the disk-like bulge in our decomposition.  

\vskip 0.25cm
{\bf NGC 1574}: BCO2007 classify this galaxy as type SB0$^-$, 
where the bar is a well-defined feature embedded in a huge extended 
disk component. The bar was also noted by Sandage $\&$ Bedke (1994) 
but was missed in RC3. De Vaucouleurs $\&$ de Vaucouleurs (1964) pointed out the
presence of a lens. Although Sandage $\&$ Bedke note that NGC 1574 
has a three-zone structure typical of S0s, BCO2007 could not identify 
a clear inner lens around the bar. 
From our $K_s$-band image it is difficult to judge for sure whether
this galaxy has a lens or not, but the surface brightness profile
shows three nearly exponential parts, which is the case also in some other  
galaxies with lenses in our sample. The bar is embedded inside the
middle exponential component of the disk. 
\vskip 0.25cm

The decomposition was carried out starting from the parameters
of the outer disk. Once reasonable approximations were found for all 
components, better solutions were searched separately for each of them.
Finally all parameters describing these components 
(except for the ellipticity of the lens) were left free
for fitting. Contrary to the usual case, for this galaxy the decomposition 
depends critically on 
the image resolution: by rebinning the image by a factor of 4
we find $B/T$ = 0.28 and $n$ = 1.5, whereas for the original image 
we obtain $B/T$ = 0.38 and $n$ = 2.9. A possible reason for the large difference 
between these two solutions is the compact 
central cusp at $r$ = 9 arcsec, found by Phillips et al. (1996): most 
probably the cusp is better resolved in the original image.
The fitted lens inside the bar region mainly improves the model
image, but does not affect the properties of the bulge.
Subtracting the bulge model 
from the original image shows a complete inner ring surrounding the 
bar, which changes the RC3 s-variety classification to r-variety.

\vskip 0.25cm

{\bf NGC 3081}: NGC 3081 is a well-known Seyfert 2 galaxy 
having four exceptionally well-defined rings (Buta $\&$ Purcell 1998): 
an R$_1$ outer ring intrinsically aligned nearly perpendicular 
to the bar, a nearly circular R$_2^{\prime}$ outer pseudoring, 
a highly elongated, aligned inner ring, and a misaligned nuclear 
ring. The deprojected radii of these features are 90, 
105, 35, and 14 arcsec, 
respectively, and color index maps (Buta 1990; Buta $\&$ Purcell 1998)
indicate that each ring is a well-defined zone of active star 
formation. The strong similarity between these rings and 
test-particle models of barred spirals
(Simkin, Su, $\&$ Schwarz 1980; Schwarz 1981) led Buta $\&$ Purcell (1998) to 
classify NGC 3081 as a resonance ring galaxy. The galaxy also 
has a complex double-barred structure detected both in the 
optical and in the near-infrared (Buta 1990; Friedli et al. 1996;
Mulchaey, Regan $\&$ Kundu 1997; Erwin 2004; Laine et al. 2002)
. The primary bar is a weak feature 
that underfills the bright inner ring. The secondary bar lies
within the nuclear ring and is misaligned with the primary 
bar by 71$^{\circ}$. Except for the R$_2^{\prime}$ outer pseudoring, 
all of these structural components of NGC 3081 are visible in 
our $K_s$-band image. The secondary bar is also surrounded by 
a lens extending to the radius of the nuclear ring.
As noted in paper II, the main bar-like feature in NGC 3081 
is the massive oval rimmed by the inner ring. The apparent 
primary bar underfills this ring/oval and seems insignificant 
in comparison. In spite of its early morphological type (S0/a), 
NGC 3081 has $J-H$, $H-K$, and $K-L'$ colors similar to those 
typically found in spiral galaxies (Alonso-Herrero et al. 1998). 
The hydrogen index also indicates a 
higher than average HI content for its type (Buta 1990). 
Star formation in the nested rings has also been studied 
by Ferruit, Wilson $\&$ Mulchaey (2000),
Buta, Byrd $\&$ Freeman (2004), and Martini et al. (2003)
although in the circumnuclear regions (210-590 pc) the stellar 
population is largely dominated by old stars (Boisson et al. 2004). 
Buta, Byrd, $\&$ Freeman (2004) show high resolution HST images of 
the inner regions of NGC 3081, revealing hundreds of barely-resolved 
star clusters in the inner ring. The distribution 
of these clusters belies the intrinsic oval shape of the ring, 
because they concentrate in arcs around the ring major axis. 
Evans et al. (1996) also noted that NGC 3081 has a bright
nucleus with an apparent bulge visible to the radius of $r$ = 10 arcsec. 

\vskip 0.25cm
In the decomposition we fitted the bulge, disk, the secondary bar/lens and the primary bar.
It was difficult to find a solution that converges, which
is not surprising taking into account the rich morphology of this galaxy. 
As always in complicated cases, a reasonable solution was found 
by repeating a series of decompositions. In the
inner regions it was not possible to fit simultaneously the bulge,
lens and the secondary bar, mainly because of the presence of the strong
nuclear ring. Instead, a solution was found were the nuclear bar and the lens were fitted
by a single function. Also, a simultaneous presence of a weak primary bar and a strong ring
leads to a solution where the bar fills the region up to the radius of the inner ring.
This somewhat overestimates the flux of the primary bar.
However, this is not expected to seriously underestimate the flux of the bulge, which
resides in the central regions of this galaxy. 
In the residual image both the inner and the nuclear ring are clearly
visible, and the secondary bar shows an ansae-type morphology.

\vskip 0.25cm
{\bf NGC 3100}: 
In the early work by Sandage $\&$ Brucato (1979) weak spiral arms were 
detected in the outer disk, although in RC3 this galaxy is classified as
a weakly barred peculiar S0.
In our analysis NGC 3100 appears as a system with a large lens and a bright 
nuclear component, to which already de Vaucouleurs $\&$  de Vaucouleurs payed attention in 1964.
Evidently, there are also weak asymmetric spiral arms in the outer disk, as demonstrated by
the image where the disk model is subtracted (Fig. 5).
The bulge component is difficult to evaluate in this galaxy. It could be
the bright nuclear component or there is no bulge at all. The larger component
fitted by a Sersic's function is clearly the large lens having $n$-parameter ($n$ = 1.8)
typical for disky systems. As we were not able to fit the small nuclear component,
no $B/T$-ratio is given in Table 6. 
This is our second candidate of galaxies that might   
be a proverbial analogue of late-type spirals among S0s. 

\vskip 0.25cm

{\bf NGC 3358}: 

The surface brightness of this early-type 
spiral (BCO type (R$_2^{\prime}$)SAB(l)a) is high and the 
main morphological features lie on an extended, almost flat 
distribution of light. The galaxy is most noteworthy for its 
strong R$_2^{\prime}$ outer pseudoring which is made 
from two bright arms that begin well off the ends of the bar 
and which partly overshoot those ends. The overshoot indicates 
a trace of an R$_1$ component. In optical images, 
the outer pseudoring is lightly patchy and is affected by 
dust lanes on the near side. The bar is a weak ansae type,
but having also some characteristics of spiral arms (see Figure 5). The bar is
embedded in a well-defined inner lens, which  
causes conflicting interpretations of the variety. There 
is trace of a secondary bar in our $K_s$-band image,
previously detected in the optical region by Buta $\&$ Crocker (1993) 
(see also Erwin 2004). The bulge seems to be the very small, 
nearly spherical component inside the secondary bar.
Unfortunately, the structure of this galaxy was too 
complicated for any reasonable decomposition.  
Because of the Freeman type II surface brightness profile, 
not even a 1D decomposition was made.

\vskip 0.25cm
{\bf NGC 3706}: This galaxy seems to have a very small, nearly edge-on inner disk, 
embedded inside a prominent bulge. The inner disk has been directly detected
in the HST image (filter = F555W) by Lauer et al.(2005) being visible above the bulge 
at least to $r$ = 1 arcsec.
They found also a stellar ring at that radius.
It is possible to fit this galaxy either by a single de Vaucouleurs type
function, or by adding an exponential inner disk.
The two-component solution gives a marginally better fit and above all,
it represents better the pre-identified structural components of this galaxy. 
In that case also the 
shape parameter of the bulge is more reasonable: $n$ = 3.6 instead of $n$ = 5.1.
Our best solution is very similar to that found by Scorza et al. (1998) in the $V$-band: 
in both cases $B/T$ = 0.92 was found (Scorza et al. 1998 
gives $D/B$ = 0.06). As in our $K_s$-band image a subtle trace of 
an inner lens is visible, we interpret the galaxy as type
E$^+$4. 

  \subsection{A comparison of 1D and 2D decomposition results}

It is generally assumed that bars in 2D decompositions do not affect the
relative masses of bulges or the properties of bulges and disks,
an understanding which is largely based on the study by de Jong (1996),
who compared 2D bulge/disk and bulge/disk/bar decompositions for spirals.
Later it was shown by Peng et al. (2002) for a few individual galaxies,
and by LSB2005 for a larger number of early-type disk galaxies, that
bars might actually play an important role in the structural decompositions.
These studies applied more advanced 2D methods than used in the study by de Jong.
In particular it was shown by LSB2005 that the omission of a bar might
overestimate the mass of the bulge even by 40$\%$. Another generally
accepted assumption is that 2D methods always give more reliable results 
than 1D methods. In the following we evaluate the reliability of this assumption.

\vskip 0.25cm

The mean values for the parameters of the bulge and the disk were 
calculated by including in the statistics all those 13 galaxies in our sample
for which it was possible
to estimate the $B/T$-ratio in a reasonable manner (excluded are NGC 3358 and NGC 3100),
as well as the 11 disk-dominated systems. The results are shown in Table 7. 
A comparison of 2D 2- and 3-component  decompositions
shows that when the primary bars or ovals are omitted, the $B/T$ ratios 
are significantly overestimated, mainly because the flux of the third component goes 
erroneously to the bulge. In the 3-component decompositions the bulges are also
more exponential than while using the 2-component method. 
However, including more components to the fit
does not affect the mean $B/T$-ratio considerably. Multiple bars/ovals
affect the $B/T$-ratio only in galaxies with very small
bulges or having high surface brightness secondary bars or ovals.  
Our best fitting multicomponent method gives $<B/T>$ = 0.25 and $<n>$ = 2.1,
which are very
similar to $<B/T>$ = 0.24 and $<n>$ = 2.1 as found earlier by LSB2005 for 
a sample of 24 Northern early-type disk galaxies.
This $B/T$ flux ratio is considerably smaller than obtained by our 2D bulge/disk
approach, which gives $<B/T>$ = 0.55. Or if the two elliptical galaxies classified as S0s 
by SGA2004 are also included, $<B/T>$= 0.61. This is very similar as the
mean $B/T$ ratio of 0.61 obtained by SGA2004 using a similar 2D bulge/disk 
decomposition approach in the same wavelength.
\vskip 0.25cm

We also find that $h_r$ is not sensitive to the 2D method used, 
indicating that a bar or an oval does not affect the 
scale length of the disk.
The 1D method generally also gives similar $h_r$, but there are
some individual cases where different $h_r$ values are obtained. The 1D method
failed to find the extended disks for the galaxies   
NGC 1079 ($h_r$ = 14 arcsec v.s. 34 arcsec) and NGC 1350 ($h_r$=47 arcsec v.s. 141 arcsec). 
For NGC 3081 the 1D method
gives larger $h_r$ (38 arcsec v.s 21 arcsec), mainly because the 1D fit
extends to a larger radial distance. Therefore
in this case the 1D method gives a more reliable estimate of $h_r$.
It is due to these exceptional cases that makes the small difference in
the mean $h_r$ obtained by the 1D and 2D methods in Table 7.
In conclusion, we did not find any evidence supporting the view that for 
multicomponent galaxies the 2D method 
automatically gives better results than the 1D method. This is 
the case only if, in addition to the bulge and the disk, at least
one more component is included to the 2D fit. 
\vskip 0.25cm

\section{The nature of bars and bulges in early-type disk galaxies}

For any hypothesis of galaxy evolution it is important to understand
whether bulges in galaxies are old classical bulges, or pseudo-bulges
formed at later phases of galaxy evolution. The main
approaches for evaluating this are the kinematic 
and photometric approaches. Bulges are expected to have
the characteristics of pseudo-bulges mainly when $n <$ 2 
(KK2004). This might be the case also when the bulge
is oriented near to the plane of the outer disk, or  
the ellipticity of the bulge is near to that
of the outer disk. However, 
the bulge is not necessarily aligned with the disk.
S0s might have many disks with different orientations,
of which NGC 1411 is a good example.
It is also assumed by KK2004 that if a galaxy has a
secondary bar or a nuclear ring inside the bulge, the bulge is  
most probably a young pseudo-bulge. Even more direct evidence
for the existence of a pseudo-bulge can be obtained kinematically:
if the ratio of the maximum velocity amplitude (V$_{max}$) 
to the mean stellar velocity dispersion ($\sigma$) 
at the same radius, V$_{max}/\sigma$, versus the ellipticity ($\epsilon$)
is above the isotropic oblate rotator curve \citep{binney1980}, the bulge
is kinematically supported. This means that stars of the bulge
have not had time to gain such large random velocities
as in the case of older classical bulges. Unfortunately 
useful kinematic data is available only for a few galaxies in our sample.

\vskip 0.25cm

We found strong evidence of pseudo-bulges in the non-elliptical
galaxies in our sample. The shape parameter of the bulge is smaller
than 2 in six of the 12 galaxies, and near to $n$=2 in ten of
the galaxies. For NGC 3358 no decomposition could be made,
but as discussed above, the bulge also for this galaxy is very small,
most probably having the characteristics of pseudo-bulges. 
Larger $n$-values for the bulge were detected only for NGC 1326 ($n$ = 3.0) 
and NGC 1574 ($n$ = 2.9), 
but even NGC 1326 has a nuclear star forming ring, and for NGC 1574
there is strong kinematic evidence showing that the 
bulge is rotationally supported. For NGC 1574 we use
$\sigma$ and  V$_{max}$ measured by Jarvis et al.(1988).
They found V$_{max}$(obs) = 40 km \ sec$^{-1}$ along the major axis, and
after correcting the inclination effect this is converted to
V$_{max}$(corr) = 145 km \ sec$^{-1}$. Based on our surface brightness profile, 
the galaxy is bulge dominated at r = 6 arcsec, 
at which radius $\sigma$ = 180 km \ sec$^{-1}$ and $\epsilon$ = 0.1 (taken from
our  $\epsilon$-profile in Fig. 2). The obtained 
V$_{max}$/$\sigma$ = 0.8, which  at $\epsilon$ = 0.1 means that
the bulge is rotationally supported. In a similar manner  
V$_{max}$/$\sigma$ were found for the galaxies NGC 1553, NGC 3100, NGC 3706
and ES0 208-G21. For NGC 3100 we used V$_{max}$(corr) = 120 km \ sec$^{-1}$
and $\sigma$ = 190 km \ sec$^{-1}$ (Carollo, Danziger $\&$ Buson 1993). The measurements
were made at the position angle of 90$^{\circ}$, which deviates only 10$^{\circ}$ from
the orientation of
the major axis of the disk. In the bulge-dominated region
at $r$ = 5-10 arcsec this means that the bulge is rotationally supported.
For NGC 1553 we use  V$_{max}$(corr) = 156 km \ sec$^{-1}$ and  $\sigma$ = 162 km \ sec$^{-1}$
at $r$ = 10 arcsec (Kormendy 1984), which also in this case means that the bulge is rotationally
supported. Quite interestingly, according to Kormendy the bright lens
dominating at larger radial distances is hotter and not rotationally supported,
indicating that {\it the lens is probably older than the bulge}. 
As expected, the elliptical galaxies NGC 3706 and ESO 208-G21 have neither kinematic nor 
photometric evidence of pseudo-bulges.
Four of the galaxies in our sample have also either 
a nuclear ring, a secondary bar or both, and generally the nuclear ring
is presently forming stars. In conclusion, all galaxies with extended disks in our sample
have some sign of a pseudo-bulge. Pseudo-bulges have been previously found in many
late-type spirals (Carollo et al. 1998), and more recently also in two early-type
spirals (Kormendy et al. 2006) with the Hubble types Sab and Sa.  
\vskip 0.25cm

An other interesting new finding of this study is that the bulges of two S0s 
in our sample are as small as in typical Sc-type spirals. For NGC 1411
$B/T$=0.08 in comparison to 0.11 found in Sc-galaxies. For NGC 3100 
the bulge is also very small, but could not be fitted in our
decompositions. Both galaxies
are dominated by lenses, NGC 1411 by a series of bright lenses, and NGC 3100 by
one dominant lens/oval. The bulge in NGC 1411 is disk-like (have small $n$), 
but none of the two galaxies have a nuclear ring, or other 
strong morphological signatures of strong star 
formation in the central regions of these galaxies. 
The bulge in NGC 1553 most probably is also extremely small, unless
the inner disk (with obvious spiral arms) is considered as a pseudo-bulge.
To our knowledge NGC 1411 and
NGC 3100 are the first candidates that might be analogs of Sc-galaxies among S0s, 
with similar $B/T$- flux ratios as typically found in Sc-galaxies. 
The existence of such galaxies has been earlier suggested in the DDO classification scheme by van den
Bergh (1976, 1998). 
These two galaxies are extreme examples of S0s that would be difficult to explain
by merging of large disk galaxies. NGC 1411 is found to be a member of the Dorado group
of 46 galaxies (Sandage 1975; Maia, da Costa $\&$ Latham 1989), which is one of the 
richest galaxy groups in the Southern hemisphere. Using a more strict definition of a group,
Garcia (1993) found the same galaxy to be a member of a small group of galaxies. NGC 1411
has also one major nearby companion, whereas NGC 3100 has many small nearby companions.
Particularly in case of NGC 1411 it is possible that a late-type spiral has transformed
to an S0c type system by stripping mechanisms.

\vskip 0.25cm

Ovals or lenses were found in 11 of the galaxies in our sample. Ovals have
minor to major axis ratios between 0.85 and 0.89, in agreement with 
the previous estimations by Kormendy (1984) for typical ovals. However, the
presence of lenses is more difficult to show, because their isophotes
are not necessarily elongated. This is also the reason why they are not  
always detected while classifying galaxies. In the decompositions 
made in this study, the lenses can be best recognized due to the 
model images (produced in the structural decompositions) that should 
have the observed galaxy morphologies.
In many cases a good agreement between the model image
and the observation was found only if a
flat inner structure was also added to the fit.  
Galaxies with prominent lenses in our sample typically have multiple exponential
parts in their surface brightness profiles.

\vskip 0.25cm
Bars, bulges and halos are related phenomena in galaxies.
It has been shown by the theoretical models (see the review by Athanassoula 2005)   
that the properties of bars are formed
in a complicated process between the halo and the disk, for example 
depending on the mass of the halo, the amount of halo mass
in the resonances, or the mass distribution of the bulge. When
these properties are favorable to angular momentum exchange
between the halo and the disk, bars are expected to evolve over time,
leading to more massive bars with larger quadrupole moments.
As bars loose angular momentum in this process these models also predict that bars
gradually slow down. However, the models are still controversial.
While the bar appears to slow down in most N-body models (Hernquist $\&$ Weinberg 1992;
Debattista $\&$ Sellwood 2000; Athanassoula 2003; Sellwood 2003; Holley-Bockelmann,
Weinberg $\&$ Katz 2005), there are some other models in which the bar slow down is barely visible 
over the Hubble time (Valenzuela $\&$ Klypin 2003). Also, because
long bars in early-type disk galaxies are repeatedly found to be fast rotating
systems (see the review by Debattista 2006) it is not clear that   
the longest bars are necessarily the most evolved bars.
In this study it is interesting to look 
at whether the properties of bars in S0s are similar or fundamentally 
different from the properties of bars in early-type spirals. 
\vskip 0.25cm

In Table 8 we compare the mean properties of primary bars in different Hubble type
bins. The properties for Sbc-Sm (T = 4-9) type galaxies are from LSB2004, 
based on a similar analysis for the galaxies in the OSUBG-survey. In order to
increase the statistics for the Sa-Sab type spirals, the data is a collection
from three samples: the OSUBGS, the sample of early-type disk galaxies 
by LSB2005, and this study. There is a clear tendency showing that
bar length increases from the late-type spirals towards the early-type spirals,
as first suggested by Elmegreen $\&$ Elmegreen (1985). The
maximum is reached for the Hubble type S0/a, but bars in S0s are clearly
shorter than bars in S0/a galaxies, although still longer than bars in late-type spirals. 
This is in agreement with the recent result obtained by Erwin (2005). 
The relative mass of the
bar, as estimated from the $A_2$ and $A_4$ Fourier amplitudes of density, has
a similar tendency. However, the bar induced maximum tangential force, $Q_g$,
decreases all the way from late-type spirals towards earlier types, reaching a 
minimum for S0s. The fact that $Q_g$ is slightly smaller in S0s is due more to the smaller
bar size and mass than to dilution by the bulge, because the
B/T ratios in S0s and S0/a galaxies are fairly similar. The way how massive bulges
dilute $Q_g$ was shown in LSB2004. 
\vskip 0.25cm

An interesting result in this study is that the properties
of bars in the spiral/S0 divide, e.g. between the S0/a and S0-galaxies 
are not exactly similar, or explainable by the dilution effects of bulges. 
Bars in S0s are shorter (1.21 v.s. 1.72, respectively) 
and less massive ($A_2$ = 0.49 v.s. 0.63) than bars in S0/a-galaxies.
The fact that $Q_g$ is slightly smaller in S0s is associated to the mass
of the bar rather than to the dilution by bulge, because the B/T-ratios
is S0s and S0/a-galaxies are fairly similar ($Q_g$= 0.24 $\pm$0.14 v.s. 0.27 $\pm$ 0.17).
The weaker bars in S0s might be due to a smaller amount of interstellar gas in 
these galaxies, which is expected to make the disks less reactive to bar formation.
Bars in S0s might be shorter and less massive also because 
they are less evolved in time (see the models by Athanassoula 2003). 
\vskip 0.25cm

\section{Summary}

A morphological analysis of a sample of 15 early-type disk galaxies,
mainly S0s, has been presented in the $K_s$-band. 
The sample is part of a more complete NIRS0S-sample of 170 S0-S0/a galaxies,
selected to have a similar size and selection criteria as in the
OSUBGS-sample for spirals.
Our main idea is to characterize the properties of bars, ovals and bulges for a 
large sample of galaxies in the spiral/S0 divide and to use this information to
evaluate the origin of S0s and how they are related to spirals.
\vskip 0.25cm 

The galaxies were analyzed by applying Fourier methods for calculating
the maximum gravitational torques, and by applying a
multicomponent 2D method for structural decompositions. 
A generalized Sersic's function is used for the bulge and exponential 
function for the disk, whereas the other non-axisymmetric components 
can be described either by a Sersic's or a Ferrers function. It was 
also possible to take into account the flattening and the shape 
parameter of the structures. For comparison 1D decomposition method
was also used. 
We found that if only the bulge and the disk are fitted, 1D and
2D methods give fairly similar results ($<B/T>$ = 0.55-0.61).

\vskip 0.25cm  

Bars in the galaxies in our sample are typically surrounded by  
lenses, and four of the galaxies have double bar/lens structures. Two of 
the galaxies classified as S0s in RC3 appeared to be elliptical galaxies (ESO 208-G21, NGC 3706), 
and three of the non-barred S0s were found to have inner disks 
(NGC 1411, NGC 3100, NGC 1553). A previously non-detected nuclear ring was discovered in 
NGC 1387, and a nuclear bar in NGC 1326. We found that bars in S0s are not
exactly similar to bars in S0/a-galaxies. They are clearly shorter and 
less massive and have slightly weaker bar torques than bars in S0/a-galaxies.  

\vskip 0.25cm 

One of the most important results 
of this study is the finding that {\it 11 non-elliptical galaxies
in our sample show strong evidence of having disk-like pseudo-bulges}:
these bulges have
surface brightness profiles that are more close to an exponential than 
to a de Vaucouleurs' type profile.
Six of the galaxies also have 
either nuclear rings, nuclear bars or nuclear disks.
All galaxies in our sample have $B/T <$ 0.4 as typically found in galaxies 
having pseudo-bulges.
We find $<B/T>$ = 0.25 and $<n>$ = 2, thus confirming the 
earlier result by LSB2005, showing that bulges in S0s are
considerably smaller and have more exponential surface brightness
profiles than assumed for their morphological types. A comparison to more simple
2D bulge/disk or 1D-decompositions shows 
the importance of the multicomponent approach: these two methods
give $<B/T>$ = 0.55-0.48, respectively, in both cases the relative mass of the bulge 
being significantly overestimated. The value $<B/T>$ = 0.55 is close to the value
0.64 obtained by SGA2004 for the
early-type disk galaxies in the same wavelength, using a similar 2D bulge/disk approach
as used in this study.
\vskip 0.25cm 

Quite interestingly, we also find two 
galaxies, NGC 1411 and NGC 3100, which due to their very small 
$B/T$ flux ratios might be proverbial analogues of Sc galaxies
among classical S0s, as discussed in the DDO galaxy classification
scheme by van den Bergh. These galaxies might be manifestations of
galaxy evolution in the Hubble sequence, possibly formed by gas stripping
mechanisms from late-type spirals. 

\section*{Acknowledgments}

EL and HS acknowledge the support of the Academy of Finland
while preparing this manuscript.
RB also acknowledges the support of NSF grant AST0507140 to the 
University of Alabama.
This research has also made use of the NASA/IPAC Extragalactic
database (NED) which is operated by the Jet Propulsion 
Laboratory, California Institute of Technology, under
contract with the National Aeronautics and Space Administration.
JHK acknowledges support from the Leverhulme Trust in
the form of a Leverhulme Research Fellowship. We also
acknowledge ESO to make possible for carrying out this 
observing run.

\clearpage

\begin{figure}
\plotone{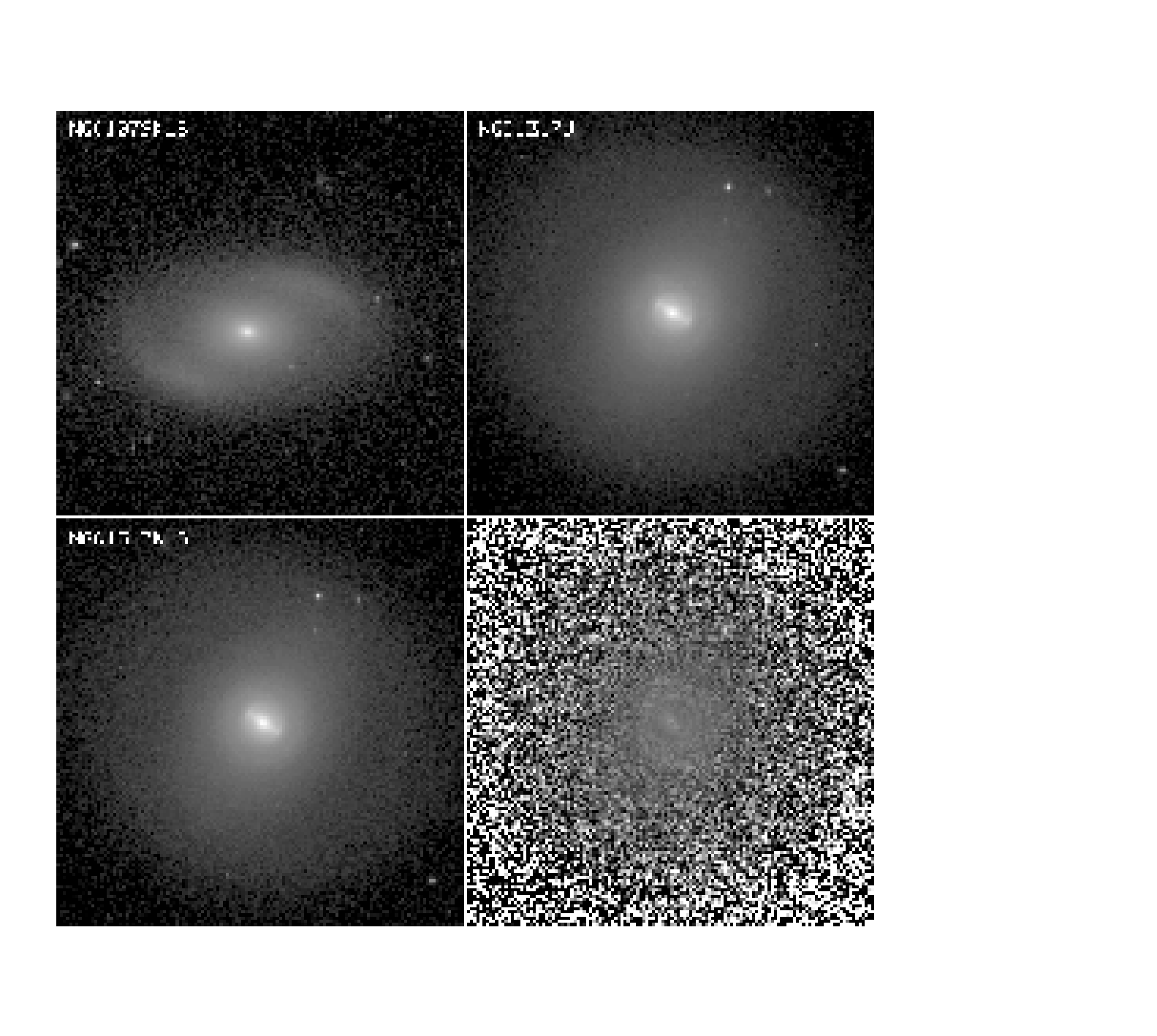}
\caption{ Direct $K_s$-band images and $J-K_s$ color index maps for the sample
galaxies. The colors index maps are coded so that redder
features are lighter. The images are shown in the plane of the sky so
that North is up and East is left. Notice that for the galaxies
NGC 1326, NGC 1512, NGC 1553, NGC 3081, NGC 3100 and NGC 3358, the
smooth in versions of the $K_s$-band images are also shown.  \label{fig1a}}
\end{figure}
\clearpage

\begin{figure}
\plotone{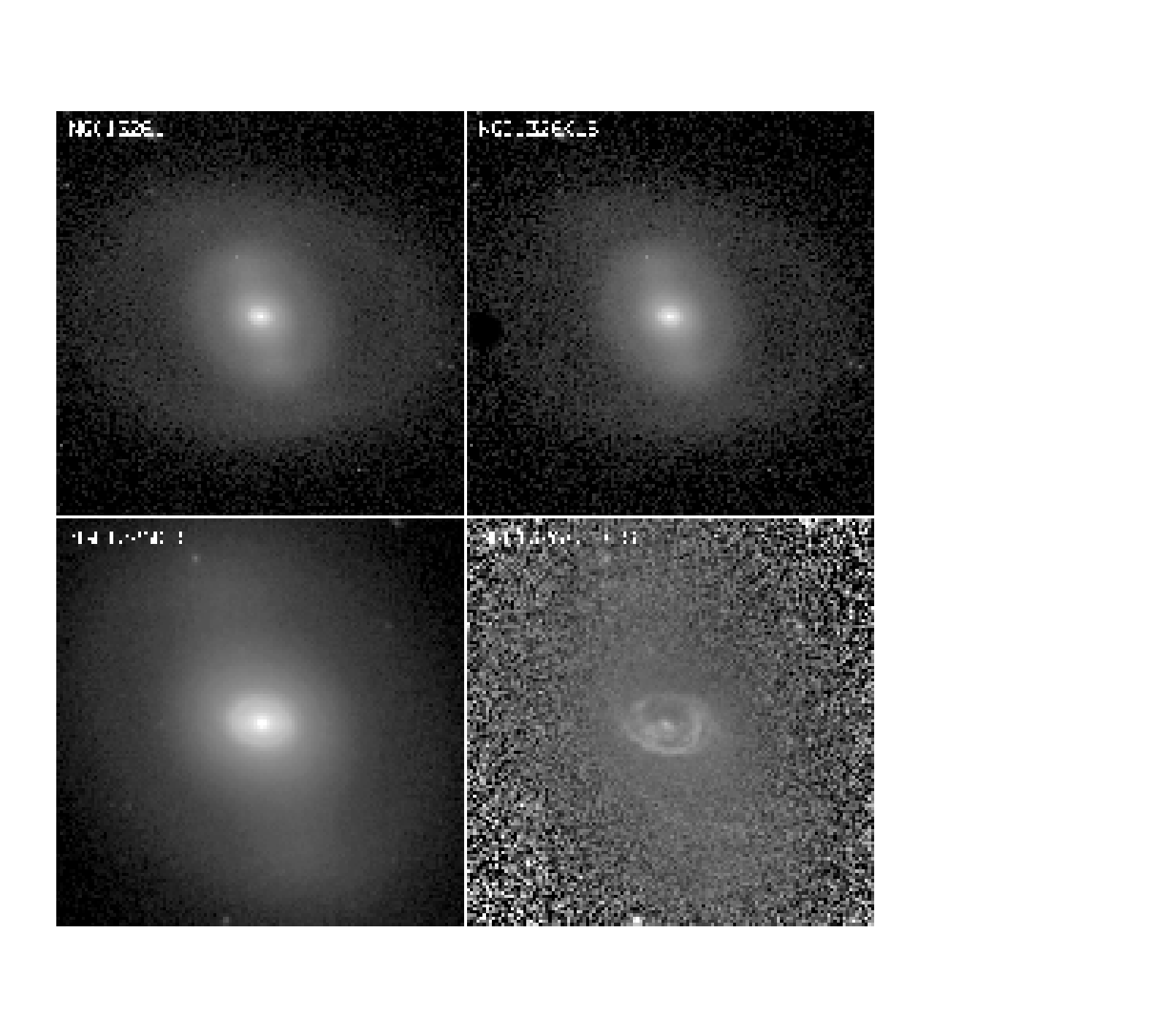} 
Figure 1 continued
\end{figure}
\clearpage

\begin{figure}
\plotone{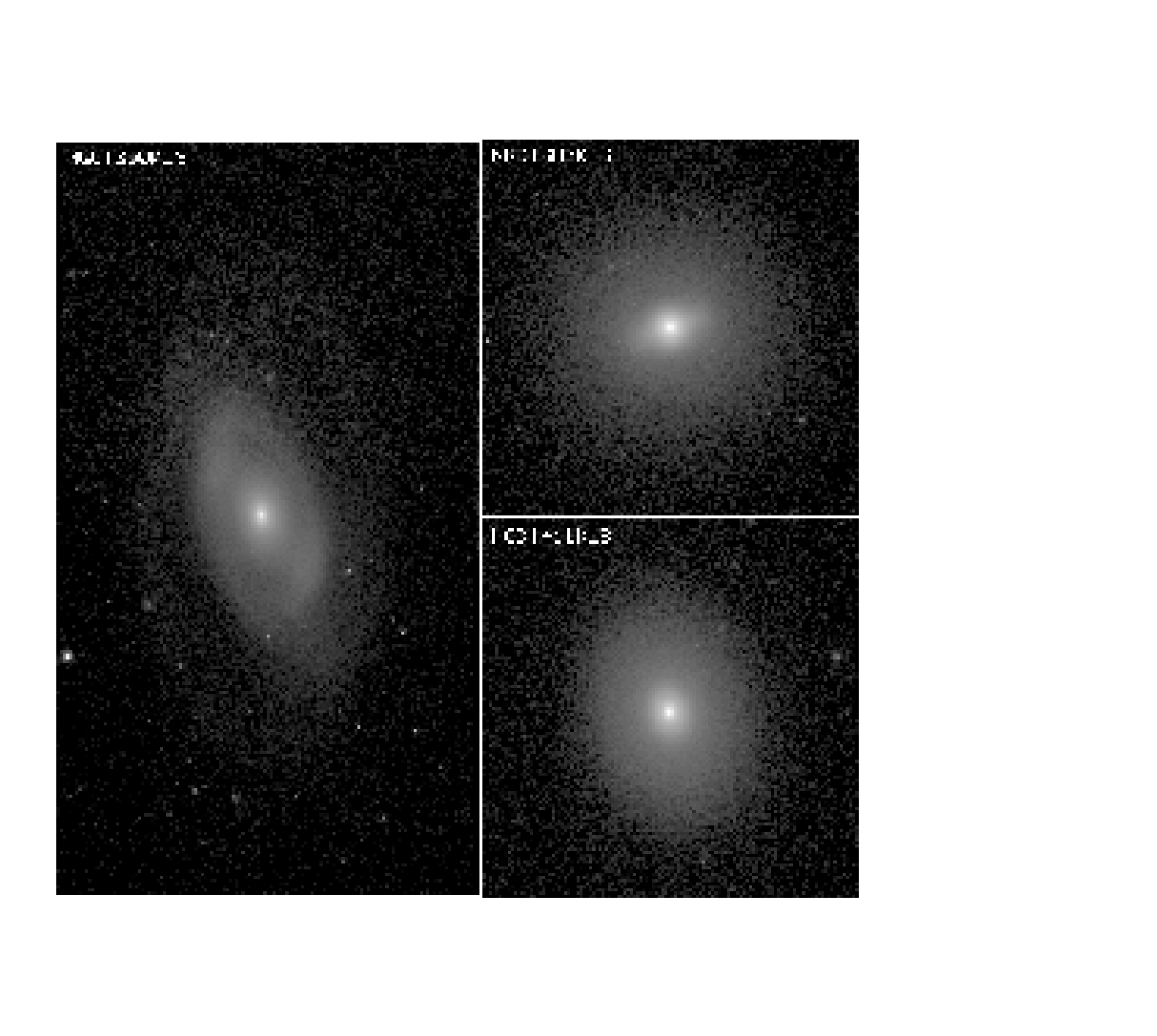} 
Figure 1 continued
\end{figure}
\clearpage

\begin{figure}
\plotone{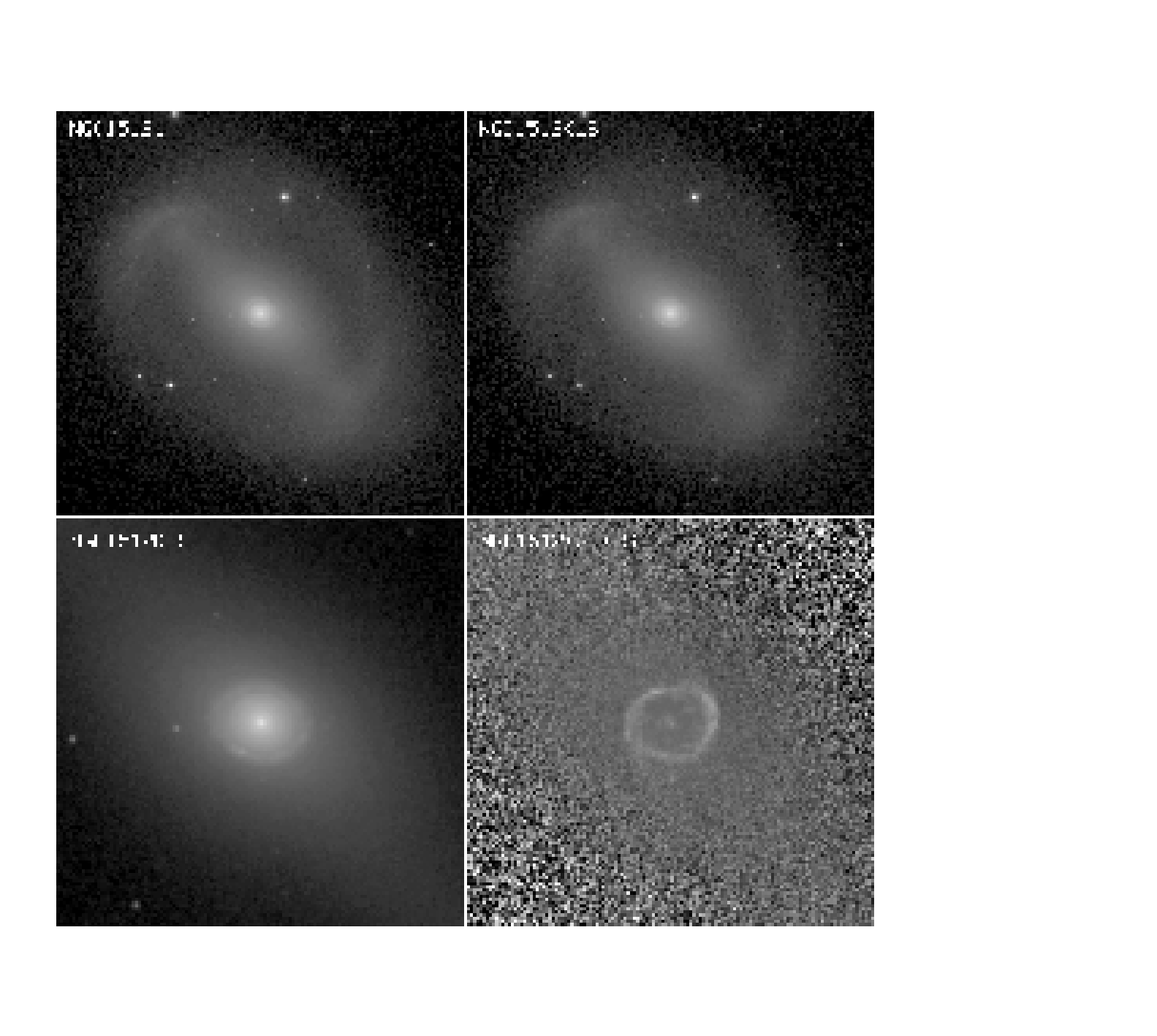}
Figure 1 continued
\end{figure}
\clearpage

\begin{figure}
\plotone{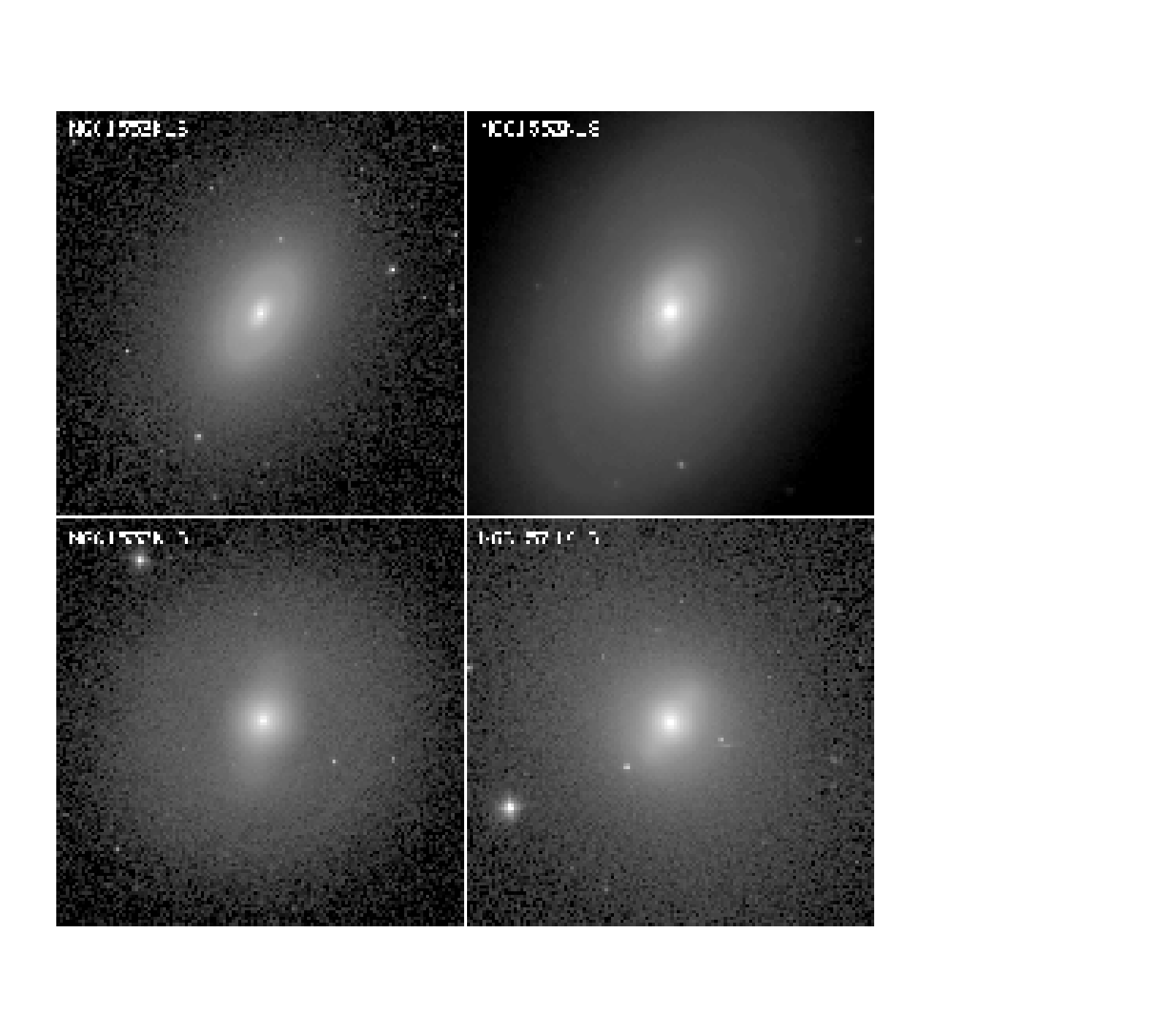}
Figure 1 continued
\end{figure}
\clearpage

\begin{figure}
\plotone{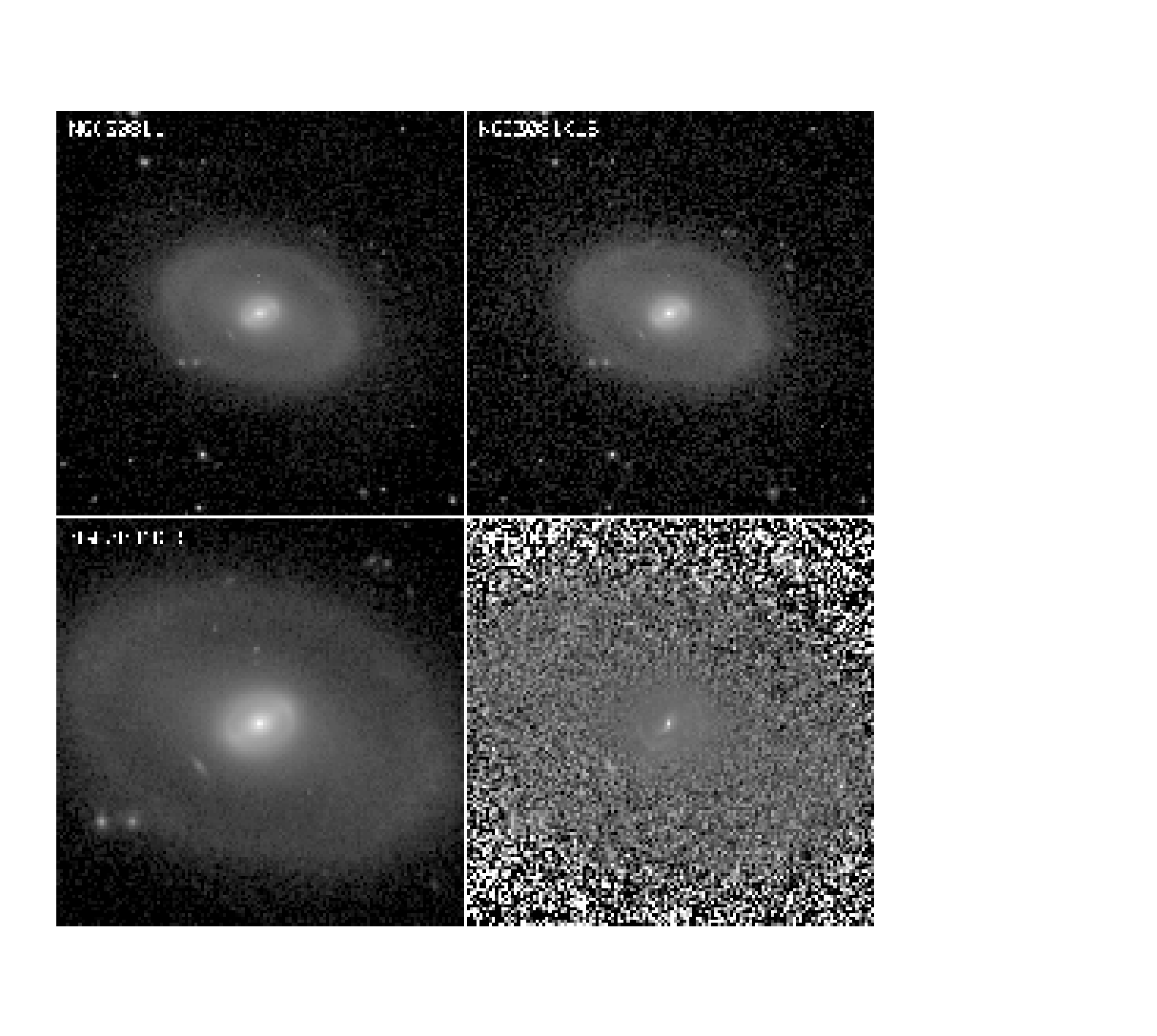}
Figure 1 continued
\end{figure}
\clearpage

\begin{figure}
\plotone{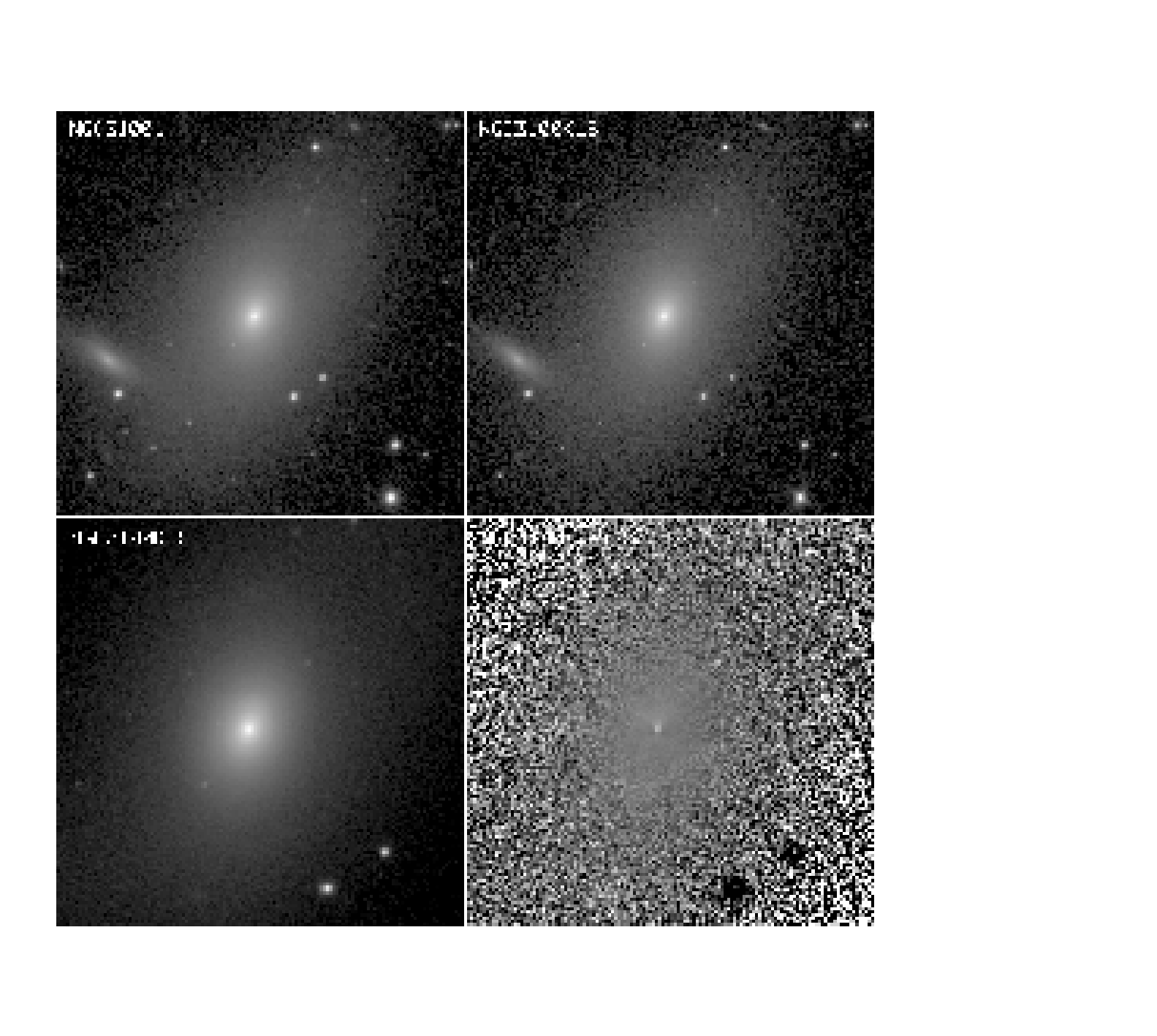}
Figure 1 continued
\end{figure}

\clearpage
\begin{figure}
\plotone{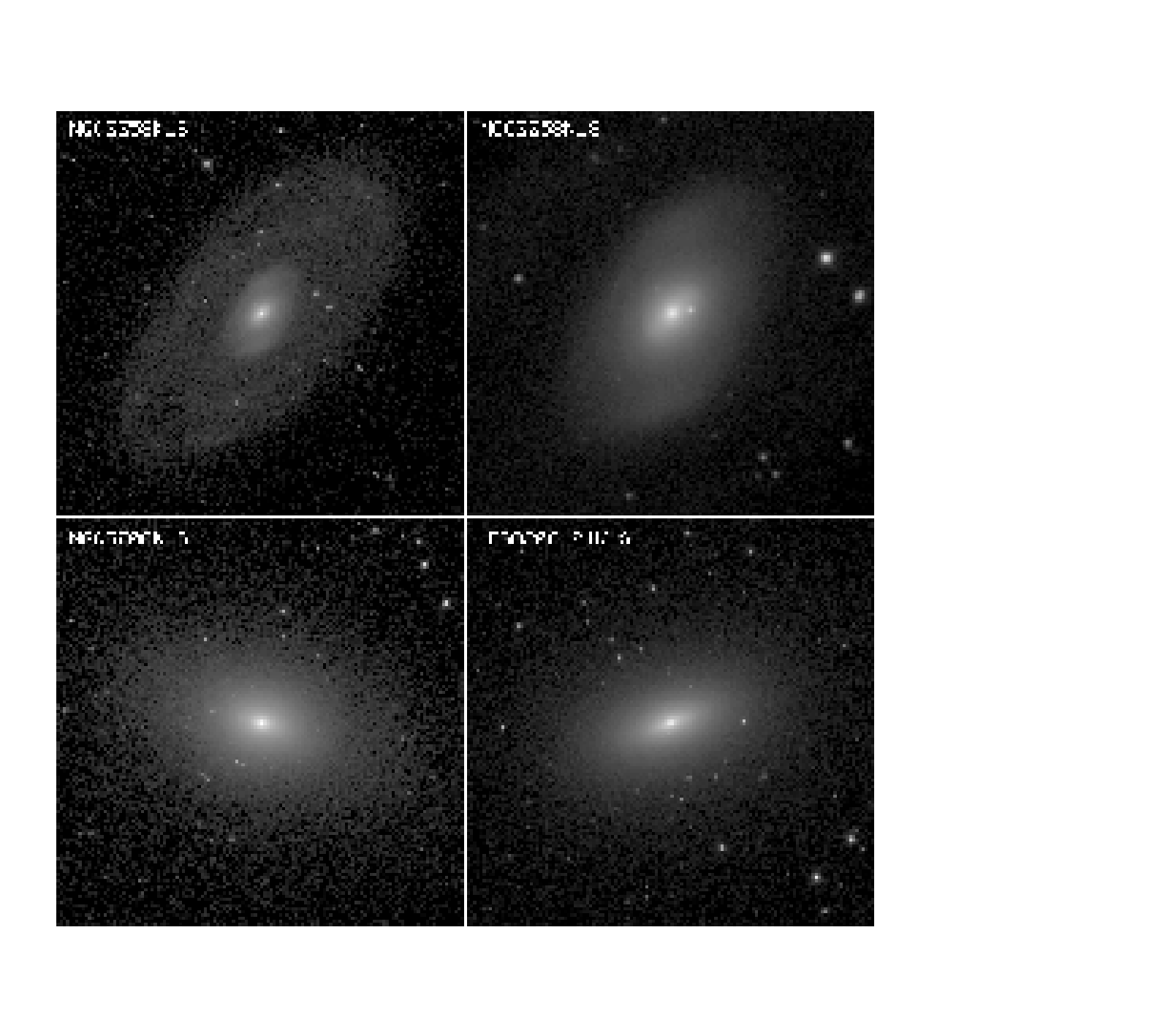}
Figure 1 continued
\end{figure}

 \clearpage

\begin{figure}
 \epsscale{1.0}
 \plotone{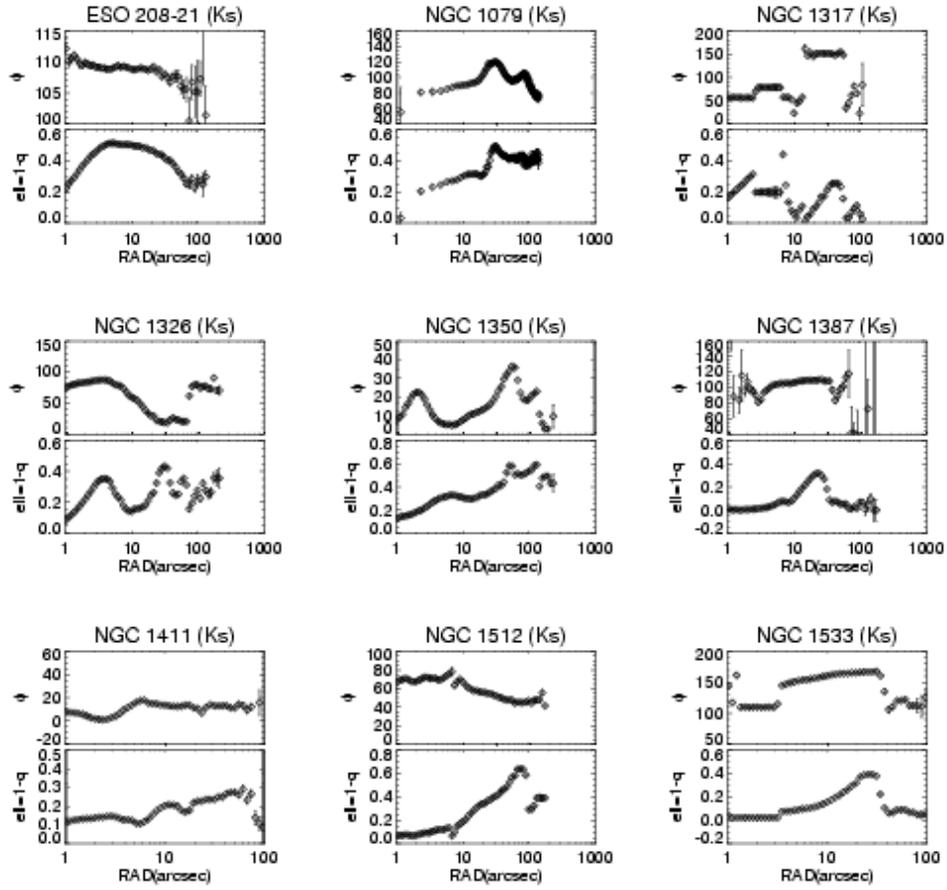}
 \caption{Radial profiles of the position angles $\phi$ and
 ellipticities $\epsilon$ = 1-$q$, where $q$ is the minor-to-major
 axis ratio, derived using the ELLIPSE routine in IRAF.   \label{fig2a}}
 \end{figure}

 \clearpage
 \begin{figure}
\epsscale{1.0}
 \plotone{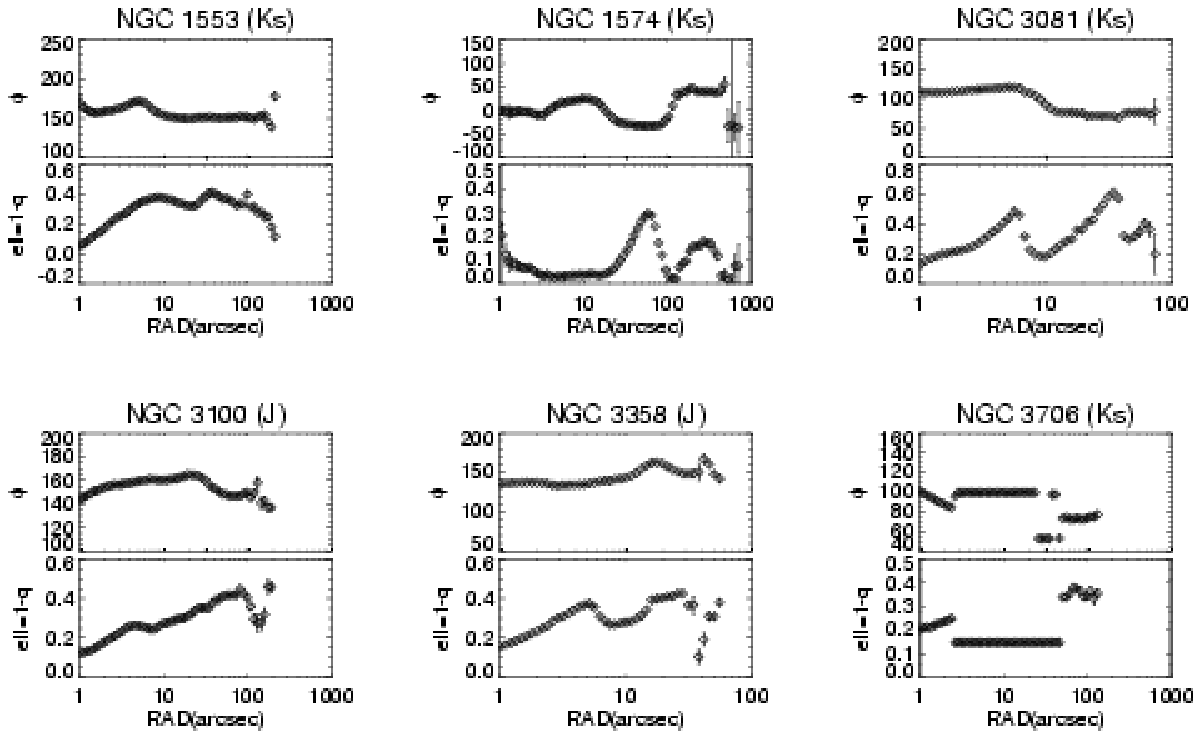}
Figure 2 continued
\caption{     \label{fig2b}}
 \end{figure}

 \clearpage

\begin{figure}
 \epsscale{1.0}
 \plotone{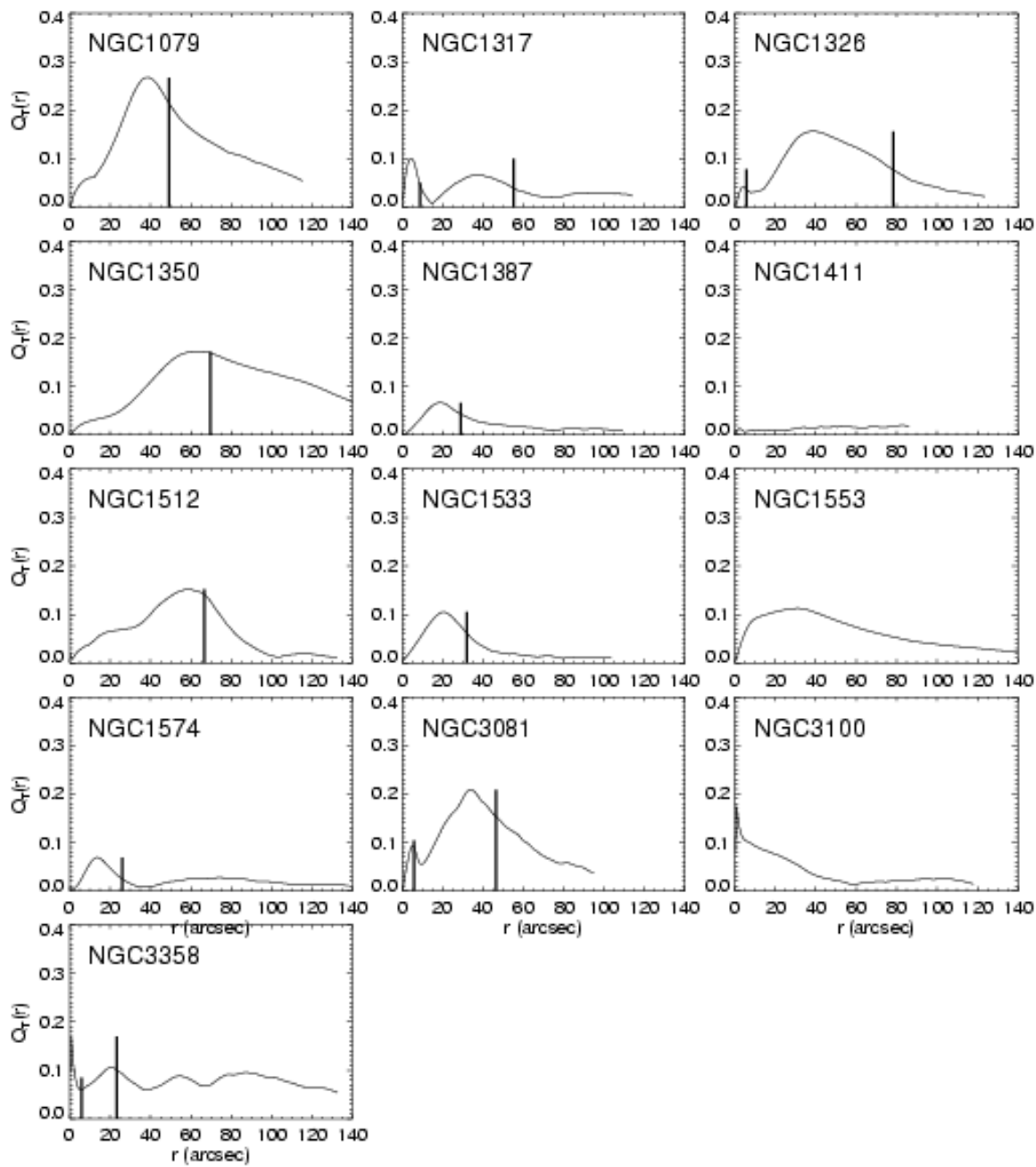}
 \caption{Radial profiles $Q_T(r)$ of the maximum relative tangential forces.
 The tangential force at each distance is normalized to the
 azimuthally averaged radial force at that distance (see in more detail in the text).
 The lengths of the primary and secondary bars are indicated by thick vertical lines.
 Notice that not all $Q_T$ maxima are induced by bars. 
 \label{fig3}}
\end{figure}

 \clearpage
 \begin{figure}
 \epsscale{.90}
 \plotone{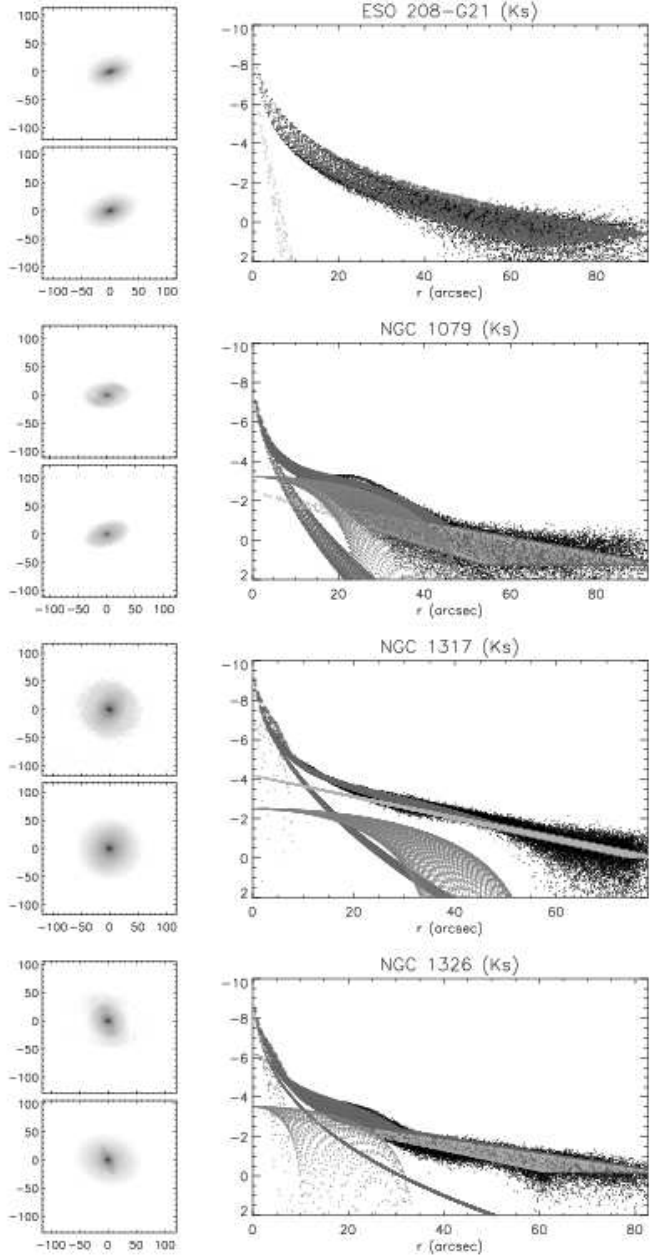}
 \caption{
 The multi-component decompositions in the $K_s$-band 
 (except for NGC 3100 in the $J$-band). 
 For each galaxy we show the observed image (upper left panel), the
 image constructed from the model functions (lower left panel), and
 the $\mu$-profile together with the fitted model functions (right panel);
$r$ gives the distance to galaxy center in sky plane. 
The observed $\mu$-profile is shown by black dots. The disk is 
 shown with {\it light grey}, bars with {\it darker grey}, and the 
 complete model and the bulge with {\it dark grey}.
 The zero-point of the surface brightness is arbitrary, but selected so that 
 one unit corresponds to one magnitude. In order to reduce the image size
only a fraction of the pixel values are shown in the plots. \label{fig4a}}
 \end{figure}

 \clearpage

 \begin{figure}
 \epsscale{.90}
 \plotone{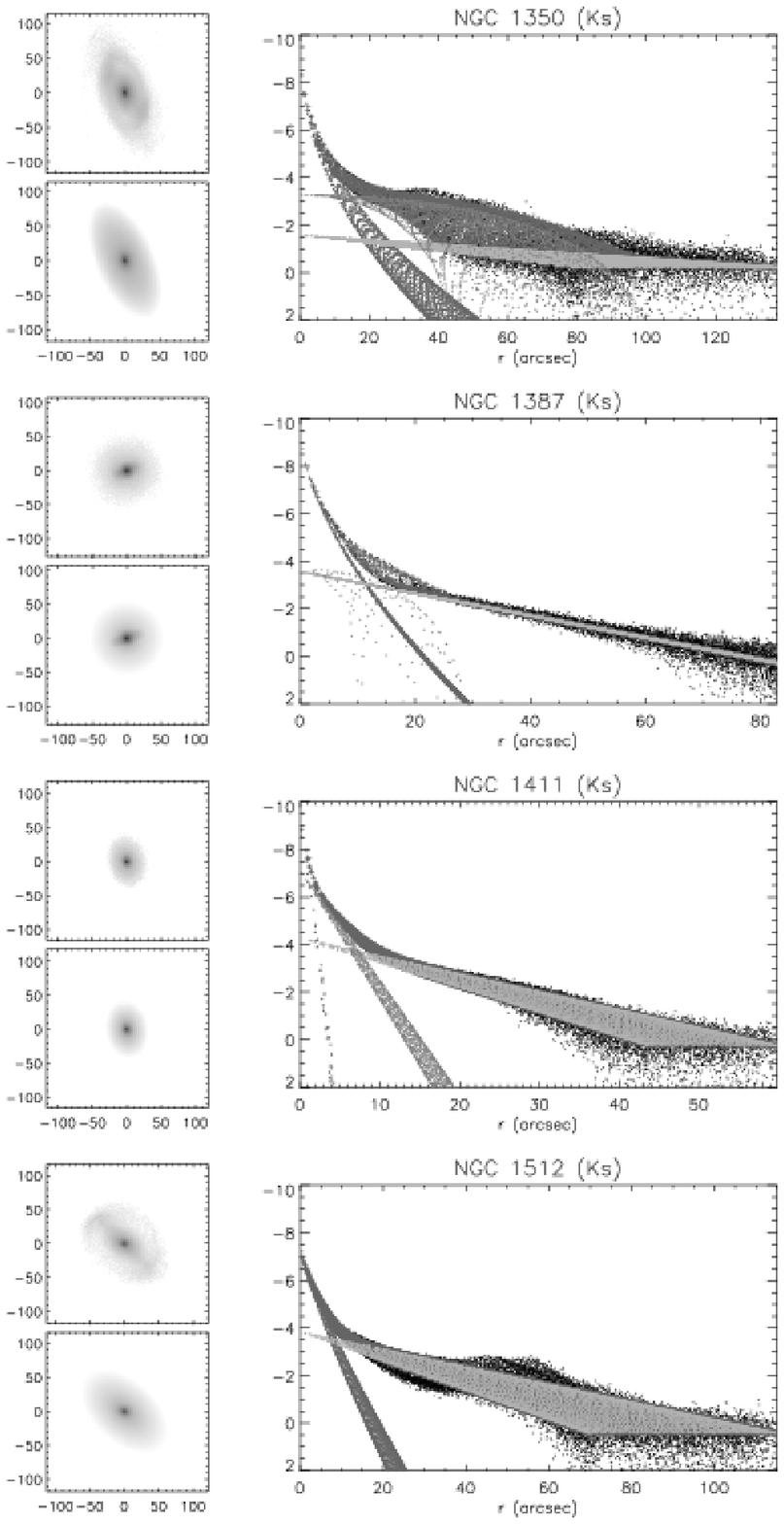}
Fig 4 continued
 \end{figure}
 \clearpage

 \begin{figure}
 \epsscale{.90}
 \plotone{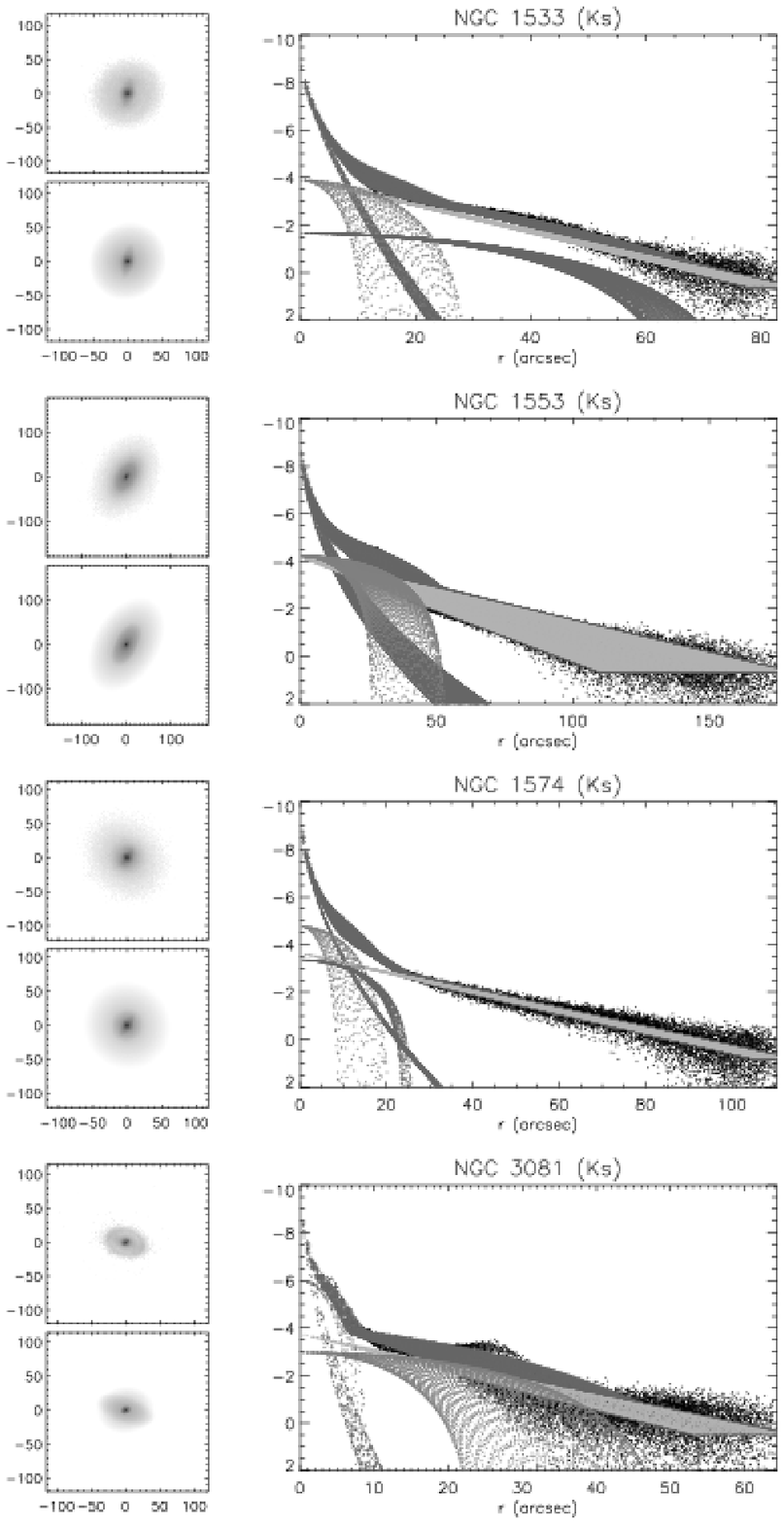}
Fig 4 continued
 \end{figure}

 \clearpage

 \begin{figure}
 \epsscale{.90}
 \plotone{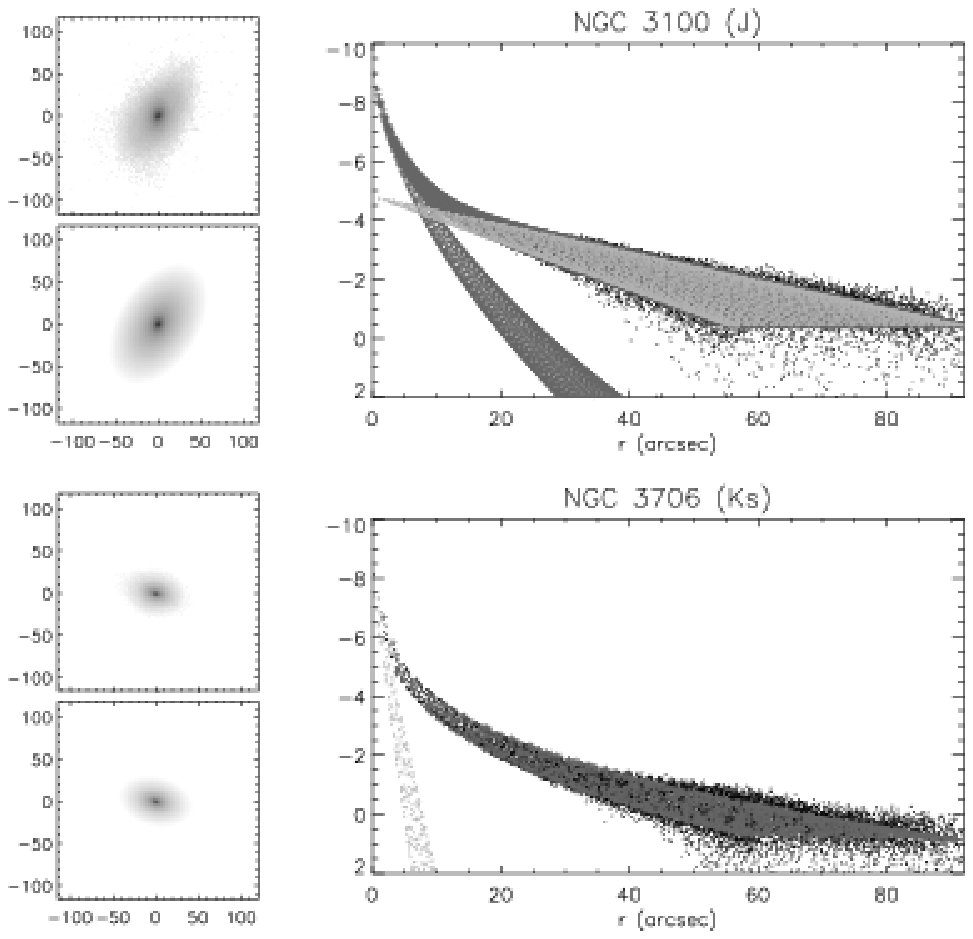}
Fig 4 continued
 \end{figure}

 \clearpage

 \begin{figure}
 \epsscale{.80}
 \plotone{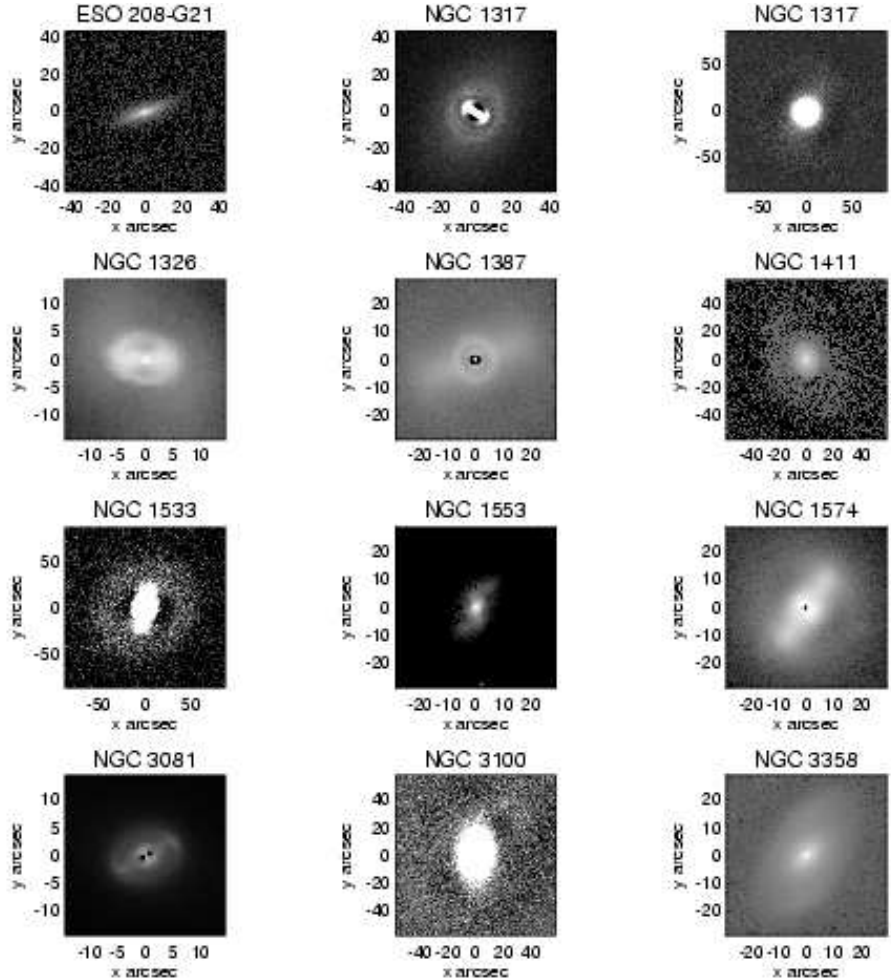}
 \caption{Residual images for nine of the galaxies in the sample.
 For ESO 208-G21 an unsharp mask-image is shown (a smoothed image
 subtracted from the original image), for the galaxies NGC 1411, NGC 1533 
 and NGC 3100, the exponential disk obtained from the decomposition 
 is subtracted from the original image, whereas for the rest of the 
 galaxies the bulge-component is subtracted. For NGC 1317 both bulge 
and disk subtracted images are shown (the upper middle and upper right 
panels, respectively).
Only the inner regions 
 of the galaxies are shown.\label{fig4}}
 \end{figure}
\clearpage

 \begin{figure}
 \epsscale{.80}
 \plotone{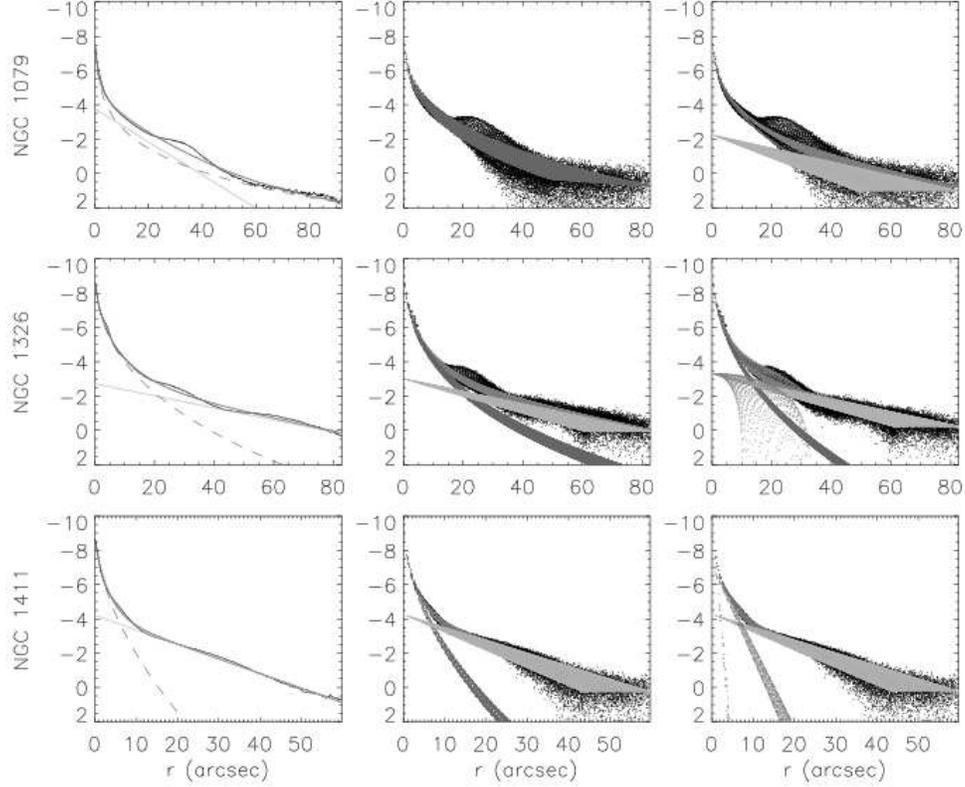}
 \caption{ 1D and 2D-decompositions are compared for the galaxies 
 NGC 1079, NGC 1326 and NGC 1411. For the last two galaxies bulge/disk and bulge/disk/bar
 2D-decompositions are shown (middle and right panels), whereas for 
 NGC 1079 we show a 2D-solution, where the whole $\mu$-profile can
 be fitted by a single Sersic's function (middle panel) or it is
 fitted by the bulge/disk alone (right panel). 
 The meaning of the colors is the same as in Fig. 4.}
 \end{figure}

\clearpage

\begin{table}
\begin{center}
\caption{The observations. If not otherwise mentioned, the classifications in column 2
are from Buta, Corwin $\&$ Odewahn (2007).  \label{tbl-1}}
\begin{tabular}{lllccc}
       &     &     &                          &               \\
\tableline\tableline
galaxy & $type_{adopted}$                     & $type_{RC3}$ & filter & FWHM & stdev \\
       &                                      &              &        & (arcsec)       \\
\tableline
ESO 208$-$G21\tablenotemark{1}  & E(d)5                                                              & SAB$^-$                  & $K_s$  & 0.63    &1.1 \\
NGC 1079                        & (R$_1$R$_2^{\prime}$)S$\underline{\rm A}$B($\underline{\rm r}$s)a  &  (R)SAB(rs)0/a        & $K_s$  & 0.84    &1.0  \\
NGC 1317\tablenotemark{2}       & (R$^{\prime}$)SAB(rl)a                                                            &SAB(r)a                & $J$    & 0.60    &1.0   \\
                                &                                                                    &                       & $K_s$  & 0.66    &2.4  \\
NGC 1326                        & (R$_1$)SAB(r)0/a                                                   &(R)SB(r)0$^+$          &  $J$   & 0.72    &1.0   \\
                                &                                                                    &                       & $K_s$  & 0.72    &1.0   \\
NGC 1350                        & (R$_1^{\prime}$)SAB(r)ab                                           &(R$^\prime$)SB(r)ab   & $K_s$   & 0.75    &1.5   \\
NGC 1387                        & SB0$^-$                                                            & SAB(s)0$^-$           & $K_s$  & 0.69    &1.1  \\
NGC 1411                        & SA(l)0$^{\circ}$                                                   & SA(r)0$^-$            & $K_s$  & 0.81    &1.2   \\
NGC 1512                        &  R$^{\prime}$)SB(r)ab pec                                          &  SB(r)a               &  $J$   & 0.69    &2.1  \\
                                &                                                                    &                       & $K_s$  & 0.63    &1.0 \\
NGC 1533                        & (RL)SB0$^{\circ}$                                                  &SB0$^-$                & $K_s$  & 0.66    &1.3  \\
NGC 1553                        & SA($\underline{\rm r}$l)0$^{+}$                                &SA(rl)0$^{\circ}$      & $K_s$  & 0.75    &1.2 \\
NGC 1574\tablenotemark{3}       & SB0$^-$                                                            & SA(s)0$^-$:           & $K_s$  & 0.75    &1.3   \\
NGC 3081                        & (R$_1$R$_2^{\prime}$)SAB(r)0/a                                     & (R)SAB(r)0/a          & $J$    & 0.69    &1.5  \\
                                &                                                                    &                       & $K_s$  & 0.63    &2.0  \\
NGC 3100\tablenotemark{1}       & SAB(s)0$^+$ pec                                                    & SAB(s)0$^{\circ}$ pec & $J$    & 0.84    &2.5  \\
NGC 3358                        & (R$_2^{\prime}$)SAB(l)a                                            &(R)SAB(s)0/a             & $K_s$  & 0.78    &0.9  \\
NGC 3706\tablenotemark{1}       & E$^+$4                                                             & SA(rs)0$^-$           & $K_s$  & 0.81    &0.8  \\

\tableline
\end{tabular}
\tablenotetext{1}{Classification in column 2 is made by Buta, based on our $K_s$-band image.}
\tablenotetext{2}{For this galaxy the residual image shows a weak classical bar.}
\tablenotetext{3}{The residual image for this galaxy shows a complete inner ring of r-variety.}
\end{center}
\end{table}

\clearpage
\begin{table}
\begin{center}
\caption{The orientation parameters.\label{tbl-2}}
\begin{tabular}{ccllcrrl}
\tableline\tableline
galaxy & filter & $\phi$ (measured) &$q$ (measured) &range       & $\phi$ (RC3)& q (RC3) & $\phi/q$ (other) \\
       &        &(degrees)          &             & (arcsec)   & (degrees)   &         & (degrees)/\\
\tableline
ESO 208-G21& $K_s$ & {\bf 109 $\pm$ 1} &{\bf 0.722 $\pm$ 0.024} & 93-116   & 110       & 0.708 & 108/0.70\tablenotemark{1}\\   
NGC 1079    & $K_s$ & {\bf 78 $\pm$ 2} &{\bf 0.590 $\pm$ 0.011} & 116-128  &  87       & 0.607 &                          \\  
NGC 1317    &  $J$  & {\bf 91 $\pm$ 2} &{\bf 0.953 $\pm$ 0.004} & 75-87    &  78       & 0.871 &   \\
NGC 1326    &  $J$  & 75.6$\pm$ 1.1        &     0.729 $\pm$ 0.000  & 134-140  &  77       & 0.741 & {\bf 73.4/0.750}\tablenotemark{5}   \\ 
NGC 1350    & $K_s$ & {\bf 6 $\pm$ 1}  &{\bf 0.586 $\pm$ 0.004} & 120-125  & 0         & 0.540 &  6/0.51\tablenotemark{2}    \\  
NGC 1387    & $K_s$ & {\bf 128 $\pm$ 10}&{\bf 0.998 $\pm$ 0.069} & 65-70    &           & 0.000 &  131/0.98\tablenotemark{8}     \\ 
NGC 1411    & $K_s$ & {\bf 9 $\pm$ 0}  &{\bf 0.729 $\pm$ 0.002} & 70-74    &  6        & 0.724 &                        \\  
NGC 1512    &  $J$  &  46  $\pm$ 0      &{\bf 0.605 $\pm$ 0.009} & 140-146  &  90       & 0.631 & 56/0.64\tablenotemark{6},{\bf 83}\tablenotemark{9}\\   
NGC 1533    & $K_s$ & {\bf 116 $\pm$ 5}&{\bf 0.942 $\pm$ 0.006} & 55-70    & 151       & 0.851 &               \\  
NGC 1553    & $K_s$ & {\bf 150 $\pm$ 6}&{\bf 0.749 $\pm$ 0.000} & 130-180  & 150       & 0.633 & /0.93 \tablenotemark{8}                       \\
NGC 1574    & $K_s$ & {\bf 31 $\pm$ 1} &{\bf 0.960 $\pm$ 0.003} & 95-100   &  35       & 0.000 &                         \\  
NGC 3081    & $J$   &  110 $\pm$ 10    &{\bf 0.861 $\pm$ 0.018} & 72-88    &  158      & 0.776 & {\bf 97/0.837}\tablenotemark{7}   \\  
NGC 3100    & $J$   & {\bf 148 $\pm$ 1}&{\bf 0.603 $\pm$ 0.012} & 105-115  & 154       & 0.513 &                      \\  
NGC 3358    & $K_s$ & {\bf 138 $\pm$ 0}&{\bf 0.581 $\pm$ 0.012} & 64-70    & 141       & 0.562 & 125/0.74\tablenotemark{3}    \\ 
NGC 3706    & $K_s$ & {\bf 76 $\pm$ 1} &{\bf 0.634 $\pm$ 0.011} & 122-140  &  78       & 0.603 &                       \\  
\tableline
\end{tabular}
\tablenotetext{1}{Martel et al. 2004, their figure 2.}
\tablenotetext{2}{Garcia-Gomez et al. 2004}
\tablenotetext{3}{Erwin 2004}
\tablenotetext{4}{Buta $\&$ Crocker 1993}
\tablenotetext{5}{Buta et al. 1998}
\tablenotetext{6}{Kuchinski et al. 2000}
\tablenotetext{7}{Buta $\&$ Purcell 1998; $\phi$ is kinematic, and $q$ is photometric}
\tablenotetext{8}{NED}
\tablenotetext{9}{Buta 1988}
\end{center}
\end{table}

\clearpage

\begin{deluxetable}{lcrcrrr}
\tablecolumns{6}
\tablewidth{0pc}
\tablecaption{The results of the Fourier analysis. \label{tbl-3}}
\tablehead{
\colhead{Galaxy} & \colhead{bar length}   & \colhead{$Q_g$}    & \colhead{$r_{Qg}$} &
\colhead{$A_2$}    & \colhead{$A_4$} \\
\colhead{} & \colhead{(arcsec)}   & \colhead{}    & \colhead{(arcsec)} & \colhead{}    & \colhead{} }
\startdata

\sidehead{Southern sample:}
NGC 1079  & 49.3     &0.268 $\pm$ 0.047 & 38.6     & 0.80       & 0.32      \\ 
NGC 1317  & 55.1     &0.065 $\pm$ 0.005 & 40.0     & 0.30       & 0.08      \\ 
NGC 1317  &  8.7     &0.100 $\pm$ 0.002 & 4.3      & 0.38       & 0.20      \\ 
NGC 1326  & 78.3     &0.157 $\pm$ 0.006 & 38.6     & 0.67       & 0.24     \\ 
NGC 1326  &  5.8     &0.040 $\pm$ 0.005 &  4.4     & 0.24       & 0.03     \\ 
NGC 1350  & 69.6     &0.172 $\pm$ 0.044 & 65.3     & 0.63       & 0.21     \\ 
NGC 1387  & 29.0     &0.067 $\pm$ 0.004 & 18.9     & 0.39       & 0.15     \\ 
NGC 1533  & 31.9     &0.105 $\pm$ 0.008 & 20.6     & 0.42       & 0.19     \\ 
NGC 1553  &          &0.113 $\pm$ 0.001 & 31.0     & 0.35       & 0.06      \\
NGC 1574  & 26.1     &0.069 $\pm$ 0.006 & 13.6     & 0.34       & 0.15     \\ 
NGC 3081  & 46.4     &0.209 $\pm$ 0.005 & 34.0     & 0.85       & 0.34     \\ 
NGC 3081  &  5.8     &0.100 $\pm$ 0.002 &  4.5     & 0.40       & 0.11     \\ 
NGC 3100  &          &0.100 $\pm$ 0.006 &  6.0     & 0.21       &              \\ 
NGC 3358  & 23.2     &0.100 $\pm$ 0.000 & 20.3     & 0.40       & 0.15     \\ 
NGC 3358  &  5.8     &0.170*$\pm$ 0.000 &  3.0     & 0.20       &            \\

\sidehead{Northern sample:}
NGC 718   & 32.2     &0.121 $\pm$ 0.011 & 18.2    & 0.41  & 0.20         \\
NGC 718   & 13.8     &0.030 $\pm$ 0.000 & 1.1     & 0.15  &                \\
NGC 936   & 59.8     &0.202 $\pm$ 0.036 & 37.0    & 0.58  & 0.30        \\
NGC 1022  & 27.6     &0.143 $\pm$ 0.009 & 15.9    & 0.41  & 0.14          \\
NGC 1400  &          &0.034 $\pm$ 0.001 &         & 0.11  &              \\
NGC 1415  & 71.3     &0.240 $\pm$ 0.046 & 47.2    & 0.80  & 0.25         \\
NGC 1440  & 27.6     &0.141 $\pm$ 0.004 & 17.7    & 0.50  & 0.23         \\ 
NGC 1452  & 46.0     &0.420 $\pm$ 0.044 & 36.1    & 0.90  & 0.58         \\
NGC 2196  &          &0.067 $\pm$ 0.018 &         & 0.15  &               \\ 
NGC 2273  & 41.4     &0.201 $\pm$ 0.013 & 20.0    & 0.57  & 0.25         \\ 
NGC 2460  &  9.2     &0.077 $\pm$ 0.028 & 4.4     & 0.14  & 0.04          \\ 
NGC 2681  & 27.6     &0.040 $\pm$ 0.000 & 16.1    & 0.23  & 0.06          \\
NGC 2681  &  4.6     &0.010 $\pm$ 0.000 & 2.3     & 0.10  & 0.03         \\
NGC 2681  & 64.4     &0.059 $\pm$ 0.001 & 54.5    & 0.30  & 0.07          \\
NGC 2781  & 62.1     &0.067 $\pm$ 0.000 & 34.5    & 0.46  & 0.10          \\ 
NGC 2855  &          &0.075 $\pm$ 0.009 &         & 0.15  &               \\ 
NGC 2859  & 69.0     &0.101 $\pm$ 0.003 & 41.6    & 0.57  & 0.22         \\ 
NGC 2859  &  6.9     &0.030 $\pm$ 0.000 & 3.5     & 0.15  & 0.03          \\ 
NGC 2911  &          &0.081 $\pm$ 0.011 &         & 0.20  &               \\ 
NGC 2983  & 41.4     &0.299 $\pm$ 0.027 & 24.6    & 0.77  & 0.45         \\ 
NGC 3626  & 39.1     &0.109 $\pm$ 0.003 & 20.5    & 0.35  & 0.08         \\ 
NGC 3941  & 29.9     &0.086 $\pm$ 0.002 & 20.0    & 0.34  & 0.14         \\ 
NGC 4245  & 50.6     &0.185 $\pm$ 0.007 & 29.2    & 0.54  & 0.20         \\ 
NGC 4340  & 78.2     &0.241 $\pm$ 0.016 & 55.9    & 0.55  & 0.27        \\ 
NGC 4340  &  4.6     &0.020 $\pm$ 0.000 & 5.8     & 0.05  &                \\
NGC 4596  & 78.2     &0.276 $\pm$ 0.050 & 48.5    & 0.67  & 0.36          \\ 
NGC 4608  & 57.5     &0.251 $\pm$ 0.011 & 37.5    & 0.77  & 0.47          \\ 
NGC 4643  & 57.5     &0.299 $\pm$ 0.007 & 40.7    & 0.85  & 0.55          \\
\enddata
\end{deluxetable}

\clearpage

\begin{table}
\begin{center}
\caption{The decomposition results: $fii_1$ is the deviation of $\phi_{bulge}$ from the orientation of the outer disk, whereas $fii_2$ 
is the deviation of $\phi_{bulge}$ from the orientation of the primary bar. fii for the other three components is the orientation
of the component in respect of the outer disk. \label{tbl-4}}
\begin{tabular}{lccccccccccccl}
\tableline\tableline 
          & bulge     &           &           &              &              &disc                   & 1                   &              &  2               &             &  3               &             &         \\
galaxy    & $n$       & $r_{eff}$ & $q$       & $fii_1$      &$fii_2$       &$h_R$                  & $q$                  & $fii$        &$q$                  & $fii$       & $q$                  &$fii$        &  $B/T$    \\
          &           & (``)      &           & ($^{\circ}$) &($^{\circ}$)  & (``)                  &                      & ($^{\circ}$) &                     & ($^{\circ}$)&                      &($^{\circ}$)  &          \\

\tableline
ESO 208-G21&4.2       &11.60      &0.60       &  0           &              &  1.4\tablenotemark{7} &                      &              &                     &             &                      &             & 0.97  \\
NGC 1079  &2.2        & 5.60      &0.73       &  7           &  65          & 34.6                  &0.51\tablenotemark{1} &  72          &                     &             &                      &             & 0.26  \\
NGC 1317  &2.1        & 5.33      &0.94       & 53           &  11          & 26.1                  &0.63\tablenotemark{1} &  64          &0.22\tablenotemark{1}& 36          &0.85\tablenotemark{2} & 36          & 0.34  \\
NGC 1326  &3.0        & 7.00      &0.99       & 21           &  15          & 36.8                  &0.28\tablenotemark{1} &  42          &0.74\tablenotemark{1}& 28          &                      &             & 0.34  \\
NGC 1350  &2.2        & 8.31      &0.71       & 13           &  57          &141.4                  &0.53\tablenotemark{6} &  43          &                     &             &                      &             & 0.25  \\
NGC 1387  &1.8        & 4.41      &0.96       & 54           &  72          & 29.6                  &0.36\tablenotemark{1} &  55          &                     &             &                      &             & 0.39  \\
NGC 1411  &1.9        & 0.72      &0.88       & 10           &   4          & 18.2                  &0.80\tablenotemark{4} &  15          &                     &             &                      &             & 0.08  \\
NGC 1512  &1.2\tablenotemark{8}   & 4.78      &0.82          & 70           &  77                   &  36.9                &              &                     &             &                      &             &        \\
NGC 1533  &1.5        & 3.88      &0.90       & 39           &  10          & 25.3                  &0.41\tablenotemark{1} &  29          &0.85\tablenotemark{3}& 29          &                      &             & 0.25  \\
NGC 1553  &1.9        & 8.51      &0.72       & 155          &              & 50.1                  &0.93\tablenotemark{8} &  100         &                     &             &                      &             & 0.21  \\ 
NGC 1574  &2.9        & 6.51      &0.97       & 31           &  24          & 34.4                  &0.38\tablenotemark{1} &   8          &0.89\tablenotemark{3}&  8          &                      &             & 0.38  \\
NGC 3081  &2.1        & 1.53      &0.68       & 19           &  60          & 21.1                  &0.51\tablenotemark{1} &  41          &0.88\tablenotemark{6}& 21          &                      &             & 0.10  \\
NGC 3100  &           &           &           &              &              & 28.7                  &0.74\tablenotemark{2} & 16           &                     &             &                      &             &      \\
NGC 3706  &3.6        &39.40      &0.67       &  1           &              &  1.4                  &                      &              &                     &             &                      &             & 0.92  \\
\tableline
\end{tabular}
\tablenotetext{1} {bar}
\tablenotetext{2} {oval fitted by Sersic function.}
\tablenotetext{3} {oval fitted by Ferrers function.}
\tablenotetext{4} {middle disk fitted by Sersic function.}
\tablenotetext{5} {bar+disk could not be fitted separately.}
\tablenotetext{6} {bar+oval could not be fitted separately.}
\tablenotetext{7} {inner disk}
\tablenotetext{8} {oval fitted by Ferrers function with $n$=1.}
\end{center}
\end{table}
\clearpage

\clearpage

\begin{table}
\begin{center}
\caption{Identification of the structural components (excluding rings, which are presented
in Table 6).\label{tbl-5}}
\begin{tabular}{ll}
              &                                                                        \\
\tableline\tableline
galaxy        & identifications                                                       \\
\tableline
ESO 208-G21  & bulge/inner disk                                                     \\ 
NGC 1079         & bulge/disk/bar/extended oval                                                      \\  
NGC 1317         & bulge/disk/$bar_1$/$bar_2$:/$oval_1$/$oval_2$   \\  
NGC 1326         & bulge/disk/$bar_1$/$bar_2$/$lens_1$/$lens_2$  \\   
NGC 1350         & flattened bulge or lens/disk/bar              \\  
NGC 1387         & bulge/disk/bar/lens                                     \\ 
NGC 1411         & bulge/middle disk/outer disk                              \\  
NGC 1512         & bulge/disk/bar/oval                        \\  
NGC 1533         & bulge/disk/bar/lens/inner ring                                      \\            
NGC 1553         & bulge/disk/lens/inner disk                                         \\
NGC 1574         & bulge/disk/bar/lens                                      \\             
NGC 3081         & bulge/disk/$bar_1$/$bar_2$/lens \\  
NGC 3100         & middle disk/outer disk                                             \\  
NGC 3358         & bulge/disk/$bar_1$/$bar_2$/oval                                     \\  
NGC 3706         & bulge/inner disk                                                    \\  
\tableline
\end{tabular}
\tablenotetext{:}{uncertain}
\end{center}
\end{table}

\clearpage

\begin{table}
\begin{center}
\caption{The major axis ring dimensions in arcseconds.\label{tbl-6}}
\begin{tabular}{lccc}
        &               &              &            \\             
\tableline\tableline
galaxy  & nuclear        & inner        & outer      \\ 

\tableline
        
NGC 1079  &               & 15        &                \\
NGC 1317  & 7, 13         & 53        &   94      \\
NGC 1326  & 6             & 28        &   71        \\
NGC 1350  &               & 80        &  170       \\ 
NGC 1387  & 6             &             &               \\
NGC 1411  &               & 37        &               \\
NGC 1512  & 8             & 64        &                 \\
NGC 1533  &               & 44        &                  \\
NGC 1574  &               & 17        &                 \\
NGC 3081  & 7             & 33        & 57              \\ 
NGC 3100  &               &           &                    \\
NGC 3358  &               &  ~22      &                  \\
\tableline
\end{tabular}
\end{center}
\end{table}

\clearpage

\begin{table}
\begin{center}
\caption{Mean $B/T$-ratio and $n_{bulge}$ as estimated using the 1D and 2D-decompositions. The errors are the errors of the mean. \label{tbl-7}}
\begin{tabular}{lccc}
                      &                 &               &               \\
\tableline\tableline
method                &  $<B/T>$        & $<n_{bulge}>$ & $<h_r>$ \\
\tableline

{\bf All (N=13):}     &                 &               &    \\

\noalign{\smallskip}

2D (final)           & 0.39 $\pm$ 0.07  & 2.4 $\pm$ 0.2 & 34.7 \\ %
2D (bulge/disk/bar)  & 0.40 $\pm$ 0.07  & 2.3 $\pm$ 0.2 & 34.7 \\ 
2D (bulge/disk)      & 0.61 $\pm$ 0.07  & 2.8 $\pm$ 0.2 & 34.7 \\ 
1D                   & 0.55 $\pm$ 0.06  & 2.9 $\pm$ 0.4 & 25.9 \\ 
\noalign{\smallskip}
\noalign{\smallskip}

{\bf Disk-dominated (N=11):}    &                 &      &        \\

\noalign{\smallskip}

2D (final)           & 0.25 $\pm$ 0.03  & 2.1 $\pm$ 0.1  & 38.5 \\ %
2D (bulge/disk/bar)  & 0.30 $\pm$ 0.03  & 2.1 $\pm$ 0.1  & 38.5 \\ 
2D (bulge/disk)      & 0.55 $\pm$ 0.06  & 2.6 $\pm$ 0.1  & 38.5 \\ 
1D                   & 0.48 $\pm$ 0.04  & 2.7 $\pm$ 0.4  & 30.2 \\ 
\tableline
\end{tabular}
\end{center}
\end{table}
\clearpage

\begin{table}
\begin{center}
\caption{Mean properties of the primary bars for different Hubble type bins. The errors are errors of the mean.  \label{tbl-8}}
\begin{tabular}{lcccc}
                 &                   &              &             &  \\             
\tableline\tableline
                 & Sbc-Sm            &Sa-Sab        & S0/a        & S0 \\ 
                 & (OSU)             & (OSU+NOT+ESO)& (NOT+ESO)   & (NOT+ESO)    \\
\tableline
        
$<Q_g>$          & 0.26 $\pm$ 0.01  &0.23 $\pm$ 0.02 & 0.18 $\pm$ 0.03 & 0.14 $\pm$ 0.02 \\
$<A_2>$          & 0.41 $\pm$ 0.02  &0.66 $\pm$ 0.06 & 0.63 $\pm$ 0.06 & 0.49 $\pm$ 0.04      \\
$<A_4>$          & 0.17 $\pm$ 0.01  &0.32 $\pm$ 0.04 & 0.29 $\pm$ 0.05 & 0.21 $\pm$ 0.03 \\
$<r_{bar}/h_r>$  & 0.91 $\pm$ 0.06  &1.42 $\pm$ 0.11 & 1.72 $\pm$ 0.15 & 1.21 $\pm$ 0.13  \\
$<r_{Qb}/h_r>$   & 0.57 $\pm$ 0.04  &0.92 $\pm$ 0.07 & 1.07 $\pm$ 0.11 & 0.80 $\pm$ 0.08  \\

$N$              & 57               & 19            & 10               & 10              \\


\tableline
\end{tabular}
\end{center}
\end{table}


\begin{thebibliography}{}


\bibitem[Alonso-Herrero et al.(1998)]{alonso1998} Alonso-Herrero, A., Simpson, C., Ward, M., Wilson, A. 1998, \apj, 495, 196
\bibitem[Athanassoula, E(2003)]{atha2003} Athanassoula, E. 2003, \mnras, 341, 1178
\bibitem[Athanassoula et al.(1990)]{atha1990} Athanassoula, E., Morin, S., Wozniak, H., Puy, D., Pierre, H., Lombard, J.,
Bosma, A. 1990, \mnras, 245, 130
\bibitem[Athanassoula(2005)]{atha2005} Athanassoula, E. 2005, Celestial Mechanics and Dynamical Astronomy 91, 9-31 (review)
\bibitem[Bekki, Warrick \& Yasuhiro(2002)]{bekki2002} Bekki, K., Warrick, J., Yasuhiro, S. 2002, \apj, 577, 651
\bibitem[Binney(1980)]{binney1980} Binney, J. 1980, \mnras, 190, 421
\bibitem[Binney \& Merrifield(1998)]{binney1998} Binney, J., Merrifield, M. 1998, ``Galactic Astronomy'' (Princeton University Press)
\bibitem[Buta(1988)]{buta1988} Buta, R. 1988, \apjs, 66, 233
\bibitem[Buta(1990)]{buta1990} Buta, R. 1990, \apj, 351, 62 
\bibitem[Buta(1995)]{buta1995} Buta, R. 1995, \apjs, 96, 39
\bibitem[Buta \& Crocker(1991)]{buta1991} Buta, R., Crocker, D. 1991, \aj, 192, 1715
\bibitem[Buta \& Crocker(1993)]{buta1993} Buta, R., Crocker, D. 1993, \aj, 105, 1344
\bibitem[Buta \& Purcell(1998)]{buta1998} Buta, R., Purcell, G. 1998, \aj, 115, 484
\bibitem[Buta et al.(1998)]{butaetal1998} Buta, R., Alpert, A., Cobb, M., Crocker, D., Purcell, G. 1998, \aj, 116, 1142
\bibitem[Buta et al.(2000)]{buta2000} Buta, R., Treuthardt, P., Byrd, G., Crocker, D. 2000, \aj, 121, 255
\bibitem[Buta, Byrd \& Freeman(2000)]{buta2004} Buta, R., Byrd, G., Freeman, T. 2004, \aj, 127, 1982
\bibitem[Buta \& Block(2001)]{buta2001} Buta, R., Block, D. 2001, \apj, 550, 243
\bibitem[Buta et al.(2006)]{buta2006} Buta, R., Laurikainen, E., Salo, H., Block, D., Knapen, J., 2006, AJ, in press (Paper II)
\bibitem[Buta, Corwin \& Odewahn(2006)]{butaetal2006} Buta, R., Corwin, H., Odewahn, S. 2006, ``The De Vaucouleurs Atlas of Galaxies'',
Cambridge, Cambridge University Press (BCO2007)
\bibitem[Boisson et al.(2004)]{boisson2004} Boisson, C., Joly, M., Pelat, D., Ward, M. 2004 \aj, 428, 373
\bibitem[Carollo, Stiavelli \& Mack(1998)]{carollo1998} Carollo, C., Stiavelli, M., Mack, J. 1998, \aj, 116, 68
\bibitem[Carollo, Danziger $\&$ Buson(1993)]{carollo1993} Carollo, C., Danziger, I., Buson, L. 1993, \mnras, 265, 553
\bibitem[Christlein \& Zabludoff(2004)]{cz2004} Christlein, D., Zabludoff, A. 2004, \apj, 616, 192 (CZ04)
\bibitem[Combes \& Sanders(1981)]{combes1981} Combes, F., Sanders, R. H. 1981, \aj, 96, 164
\bibitem[Corwin, de Vaucouleurs \& de Vaucouleurs(1985)]{corwin1985} Corwin, H. G., de Vaucouleurs, A., de Vaucouleurs, G. in 
``Southern Galaxy Catalogue of 5481 Galaxies South of DEclination 17 Degrees found in 1.2 M U.K. Smidth IIIA-J Plates,
Austin, Texas, 185
\bibitem[Crocker, Baugus \& Buta(1996)]{crocker1996} Crocker, D., Baugus, P., Buta, R. 1996 \apjs, 105, 353
\bibitem[Debattista(2006)]{debattista2006} Debattista, V. 2006, astro-ph/0601277
\bibitem[Debattista \& Sellwood(2000)]{debattista2000} Debattista, V., Sellwood J.A. 2000, \apj, 543, 704 
\bibitem[de Grijs et al.(2003)]{degrijs2003} de Grijs, R., Anders, P., Bastian, N., Lynds, R., Lamers, H., O'Neil, E. 2003, \mnras, 343, 1285
\bibitem[de Jong(1996)]{dejong1996} de Jong, R., 1996, \aap, 313, 45
\bibitem[de Vaucouleurs \& de Vaucouleurs(1964)]{devauc1964} de Vaucouleurs, G., De Vaucouleurs, A. 1964, ``Reference Catalogue of Bright Galaxies'' (Austin: University of Texas) 
\bibitem[de Vaucouleurs(1974]{decauc1974} de Vaucouleurs, G. 1974, in Formation and Dynamics of Galaxies,
IAU Symposium No. 58, J. R. Shakeshaft, ed., Dordrecht, Reidel, p. 1.
\bibitem[de Vaucouleurs et al.(1991)]{devauc1991} de Vaucouleurs, G., de Vaucouleurs, A., Corwin, H., Buta, R. J., Paturel, G., Fouqu\'e, P. 1991, ``Third Reference Catalog of Bright Galaxies'' (New York: Springer) (RC3)
\bibitem[de Souza, Gadotti \& dos Anjos(2004)]{souza2004} de Souza, R., Gadotti, A., dos Anjos, S. 2004, \apjs, 153, 411 (SGA2004)
\bibitem[Erwin(2004)]{erwin2004} Erwin, P. 2004, \aap, 415, 941
\bibitem[Erwin(2005)]{erwin2005} Erwin, P. 2005, \mnras, 364, 283
\bibitem[Eskridge et al.(2002)]{esk2002} Eskridge, P. et al. 2002, \apjs , 143, 73 
\bibitem[Evans et al.(1996)]{evans1996} Evans, I., Koratkar, A., Storchi-Bergmann, T., Kirkpatrick, H., Heckman, T, 
Wilson, A. 1996, \apjs, 105, 93
\bibitem[Ferruit, Wilson \& Mulchaey(2000)]{ferruit2000} Ferruit, P., Wilson, A., Mulchaey, J. 2000, \apjs, 128, 139
\bibitem[Fernandes, Storchi-Bergmann \& Schmitt(1998)]{fernandes1998} Fernandes, C., Storchi-Bergmann, T., Schmitt, H. 1998, \mnras, 297, 579
\bibitem[Friedli et al.(1996)]{friedli1996} Friedli, D., Wozniak, H., Rieke, M., Martinet, L., Bratschi, P. 1996, \aaps, 118, 461
\bibitem[Garcia(1993)]{garcia1993} Garcia, A. M. 1993, \apjs, 100, 47
\bibitem[Garcia-Barreto et al.(1991)]{garcia1991} Garcia-Barreto, J., Dettmar, R., Combes, F., Gerin, M., Koribalski, B. 1991, \rmxaa, 22, 197
\bibitem[Garcia-Gomez et al.(2004)]{garcia2004} Garcia-Gomez, C., Barbera, C., Athanassoula, E., Bosma, A., Whyte, L. 2004, \aap, 421, 595
\bibitem[Hernquist \& Weinberg(1992)]{hernquist1991} Hernquist, L., Weinberg, M. 1992, \apj, 446, 717
\bibitem[Hawarden et al.(1979)]{hawarden1979} Hawarden, T., van Woerden, H., Goss, W., Mebold, U., Peterson, B. 1979, \aaps, 76, 230
\bibitem[Holley-Bockelmann, Weinberg \& Katz(2005)]{holley2005} Holley-Bockelmann, K., Weinberg, M., Katz, N. 2005, \mnras, 363, 991
\bibitem[Jarvis et al.(1988)]{jarvis1988} Jarvis, B., Dubath, P., Martinet, L., Bacon, R. 1988, \aaps, 74, 513 
\bibitem[Jungwiert, Combes \& Axon(1997)]{jung1997} Jungwiert, B., Combes, F., Axon, D. 1997, \aaps, 125, 479
\bibitem[Kennicutt(2003)]{kennicutt1993} Kennicutt, R. C. Jr. et al. 2003, \pasp, 115, 928
\bibitem[Kent(1985)]{kent1985} Kent, S. 1985, \apjs, 59, 115 
\bibitem[Knapen et al.(2006)]{knapen2006} Knapen, J. H., Mazzuca, L. M., B{\"o}ker, T., Shlosman, I., Colina, L., Combes, F., \& Axon,
D. J.\ 2006, \aap, 448, 489
\bibitem[Kormendy(1979)]{kormendy1979} Kormendy, J., 1979, \apj, 227, 714
\bibitem[Kormendy(1984)]{kormendy1984} Kormendy, J., 1984, \apj, 286, 116
\bibitem[Kormendy \& Bender(1996)]{kormendy1996} Kormendy, J., Bender, R. 1996, \apj, L119
\bibitem[Kormendy \& Kennicutt(2004)]{kk2004} Kormendy, J., Kennicutt, R. Jr. 2004, Ann Rev. Astr. Ap., Vol. 42, 603 (KK2004) 111
\bibitem[Kormendy  et al.(2006)]{kormrndy2006} Kormendy, J, Cornell,, M. E., Block, D., Knapen, J., Allard, E. 2006, \apj, 642, 765
\bibitem[Kuchinski et al.(2000)]{kuchinski2000} Kuchinski, L., Freedman, W., Madore, B., Trewhella, M., Bohlin, R., 
Cornett, R., Fanelli, M., Marcum, P., Neff, S., O'Connell, R. and 4 coauthors 2000, \apjs, 131, 441
\bibitem[Laine et al.(2002)]{laine2002} Laine, S., Shlosman, I., Knapen, J. H., \& Peletier, R. F.\ 2002, \apj, 567, 97
\bibitem[Laurikainen \& Salo(2000)]{lauri2000} Laurikainen, E., Salo, H. 2000, \aaps, 141, 103
\bibitem[Laurikainen \& Salo(2002)]{lauri2002} Laurikainen, E., Salo, H., 2002 \mnras, 337, 1118
\bibitem[Laurikainen et al.(2004)]{laurietal2004} Laurikainen, E., Salo, H., Buta, R., Vasylyev S., 2004, \mnras, 355, 1251 
\bibitem[Laurikainen, Salo \& Buta(2004)]{lauri2004} Laurikainen, E., Salo, H., Buta, R., 2004, \apj, 607, 103 (LSB2004)
\bibitem[Laurikainen, Salo \& Buta(2005)]{lauri2005} Laurikainen, E., Salo, H., Buta, R. 2005, \mnras, 362, 1319 (LSB2005, Paper I)
\bibitem[Lauer et al.(2005)]{lauer2005} Lauer, T., Faber, S., Gebhardt, K., Richstone, D., Tremaine, S., Ajhar, E., Aller, M., Bender, R., 
Dressler, A., Filippenko, A., and 7 coauthors 2005, \aj, 129, 2138
\bibitem[Maia(1989)]{maia1989} Maia, M. A., da Costa, L. N., Latham, D. W. 1989, \apjs, 69, 809
\bibitem[Martel et al.(2004)]{martel2004} Martel, A., Ford, H., Bradley, L., Tran, H., Menanteau, F., Tsvetanov, Z., Illingworth, G., 
Hartig, G., Clampin, M. 2004, \aj, 128, 2758
\bibitem[Maoz et al.(2001)]{maoz2001} Maoz, D., Barth, A., Ho, L., Sternberg, A., Filippenko, A. 2001, \aj 121, 3048
\bibitem[Martini et el.(2003)]{martini2003} Martini, P., Regan, M., Mulchaey, J., Pogge, R. 2003, \apjs, 146, 353
\bibitem[Mulchaey, Regan \& Kundu(1997)]{mulchaey1997} Mulchaey, J., Regan, M., Kundu, A. 1997, \apjs, 110, 299
\bibitem[Pahre(1999)]{pahre1999} Pahre, M. 1999, \apjs, 124, 127
\bibitem[Papovich et al.(2003)]{papovich2003} Papovich, C., Giavalisco, M., Dickinson, M., Conselice, C., Ferguson, H. 2003, \apj, 598, 827
\bibitem[Peng et al.(2002]{peng2002} Peng, C., Ho, L., Impey, C., Rix, H. 2002, \aj, 124, 266
\bibitem[Pfenniger(1984)]{pfenniger1984} Pfenniger, D. 1984, \aap, 134, 373
\bibitem[Phillips et al.(1996)]{phillips1996} Phillips, A., Illingworth, G., MacKenty, J., Franx, M. 1996, \aj, 111, 1566
\bibitem[Quillen, Frogel \& Gonzalez(1994)]{quillen1994} Quillen, A., Frogel, J., Gonzalez, R. 1994, \apj, 437, 162
\bibitem[Raha et al.(1991)]{raha1991} Raha, N., Sellwood, J., James, R., Kahn, F. 1991, Nature, 352, 411
\bibitem[Rampazzo(1988)]{rampazzo1988} Rampazzo, R. 1988, \aap, 204, 81
\bibitem[Rautiainen et al.(1999)]{rautiainen1999} Salo, H., Rautiainen, P., Buta, R., Purcell, G., Cobb, M., 
Crocker, D, Laurikainen, E. 1999, \aj, 117, 792
\bibitem[Salo, Laurikainen \& Buta(2004)]{salo2004} Salo, H., Laurikainen, E., Buta, R. 2004, in ``Penetrating Bars Through Masks of Cosmic Dust'',
eds. D. Block, I. Puerari, K. Freeman, R. Groess, E. Block, (Springer), p. 673
\bibitem[Sandage(1975)]{sandage1975} Sandage, A. 1975, \apj, 202, 56
\bibitem[Sandage \& Bedke(1994)]{sandage1994} Sandage, A., Bedke J. 1994, ``The Carnegie Atlas of Galaxies'' (Washington: Carnegie Inst.)(CAG)
\bibitem[Sandage \& Brucato(1979)]{sandage1979} Sandage, A., Brucato R. 1979, \aj, 84, 472
\bibitem[Schweizer(1980)]{schweizer1980} Schweizer, F. 1980, \apj, 237, 303
\bibitem[Schwarz(1981)]{schwarz1981} Schwarz, M. 1981, \apj, 247, 77
\bibitem[Scorza et al.(1998)]{scorza1998} Scorza, C., Bender, R., Winkelmann, C., Capaccioli, M., Macchetto, D. F. 1998, \aaps, 131, 265
\bibitem[Sellwood(2003)]{sellwood2003} Sellwood, J.A. 2003, \apj, 587, 638
\bibitem[Simkin, Su \& Schwarz(1980]{simkin1980} Simkin, S., Su, H., Schwarz, M. 1980, \apj, 237, 404
\bibitem[Storchi-Bergman et al.(1996)]{storchi1996} Storchi-Bergmann, T., Rodriguez-Ardila, A. Schmitt, H., Wilson, A., Baldwin, J. 1996, \apj, 472, 83
\bibitem[Valenzuela \& Klypin(2003)]{valenzuela2003} Valenzuela, O., Klypin, A. 2003, \mnras, 345, 406
\bibitem[van den Bergh(1976)]{vandenberg1976} van den Bergh, S. 1979, \apj, 206, 883
\bibitem[van den Bergh(1998)]{vandenberg1998} van den Bergh, S. 1998, ``Galaxy Morphology and Classification'' (Cambridge University Press)
\bibitem[Wozniak et al.(1995)]{wozniack1995} Wozniak, H., Friedli, D., Martinet, P., Bratschi, P. 1995, \aap, 111, 115

\end{thebibliography}
\end{document}